\documentclass[journal]{IEEEtran}
\usepackage{xfrac}
\usepackage{amsmath}
\usepackage{cases}
\usepackage{amssymb,mathrsfs,dsfont}
\usepackage{enumerate}
\usepackage{color}
\usepackage{soul}
\usepackage{textcomp}
\usepackage{tikz}
\usepackage{pgfplots}
\usepackage{mathtools}
\usepackage{scrextend}
\usetikzlibrary{dsp,chains}
\mathtoolsset{showonlyrefs=true}
\usetikzlibrary{shapes,arrows}
\usetikzlibrary{positioning,shapes,shadows,arrows,scopes,decorations}
\usetikzlibrary{shapes.multipart}
\usetikzlibrary{snakes}
\newtheorem{theorem}{Theorem}
\newtheorem{lemma}{Lemma}
\newtheorem{proposition}{Proposition}

\newtheorem{corollary}{Corollary}
\newtheorem{remark}{Remark}

\def\md{\mathbb}

\def\eps{\varepsilon}
\def\tn{\textnormal}
\def\wt{\widetilde}
\def\wh{\widehat}
\def\RealF{\md{R}}

\def\Expt{\md{E}}

\def\db{\mathrm{dB}}
\def\latticeN{{N_\Lambda}}
\def\SKn{k}
\def\SKN{K}
\def\Ncode{{N_\Lambda}}
\def\V0{\mathcal{V}_0}

\def\dotleq{\stackrel{.}{\leq}}
\def\dotgeq{\stackrel{.}{\geq}}

\def\maxerr{\text{maxerr}}

\newcommand{\bM}[1]{\boldsymbol{#1}}

\newcommand{\dfn}{ \stackrel{\tn{def}}{=} }

\newcommand{\moduloOp}[1] {\mathbb{M}_d\left[#1\right]}
\newcommand{\moduloLattice}[1] {\mathbb{M}_{\Lambda}\left[#1\right]}
\newcommand{\quantLattice}[1] {\mathbb{Q}_{\Lambda}\left[#1\right]}
\newcommand{\qfunc}[1] {Q\left(#1\right)}
\newcommand{\snr}{\mathrm{SNR}}
\newcommand{\bsnr}{\mathrm{S}\wt{\mathrm{N}}\mathrm{R}}
\newcommand{\dsnr}{\Delta\snr}
\newcommand{\Pe}{p_e}
\newcommand{\PeDMC}{\delta}
\newcommand{\Pet}{\delta}
\newcommand{\Pmod}{p_\textrm{mod}}
\newcommand{\Pdec}{p_\textrm{dec}}
\newcommand{\latticeLoose}{L}

\IEEEoverridecommandlockouts

\renewcommand{\baselinestretch}{1.15}
\newcommand{\bs}[1]{\boldsymbol{#1}}
\usepackage[noadjust]{cite}

\def\psic{{\Psi_4}}
\def\psicdb{\Psi_{4,\db}}
\def\psid{\Psi_3}
\def\psiddb{\Psi_{3,\db}}

\begin{document}

\title{Interactive Schemes for the AWGN Channel with Noisy Feedback}

\author{Assaf~Ben-Yishai and Ofer~Shayevitz\thanks{The authors are with the Department of EE--Systems, Tel Aviv University, Tel Aviv, Israel \{assafbster@gmail.com, ofersha@eng.tau.ac.il\}. The work of A. Ben-Yishai was partially supported by an ISF grant no. 1367/14. The work of O. Shayevitz was supported by an ERC grant no. 639573, a CIG grant no. 631983, and an ISF grant no. 1367/14. This paper was presented in part at Allerton 2014 and ISIT 2015.
}}

\maketitle

\begin{abstract}
We study the problem of communication over an additive white Gaussian noise (AWGN) channel with an AWGN feedback channel. When the feedback channel is noiseless, the classic Schalkwijk-Kailath (S-K) scheme is known to achieve capacity in a simple sequential fashion, while attaining reliability superior to non-feedback schemes. In this work, we show how simplicity and reliability can be attained even when the feedback is noisy, provided that the feedback channel is sufficiently better than the feedforward channel. Specifically, we introduce a low-complexity low-delay interactive scheme that operates close to capacity for a fixed bit error probability (e.g. $10^{-6}$). We then build on this scheme to provide two asymptotic constructions, one based on high dimensional lattices, and the other based on concatenated coding, that admit an error exponent significantly exceeding the best possible non-feedback exponent. Our approach is based on the interpretation of feedback transmission as a side-information problem, and employs an interactive modulo-lattice solution. 
\end{abstract}

\section{Introduction}\label{sec:introduction}
While feedback cannot increase the capacity of point-to-point memoryless channels \cite{ShannonFeedback}, there exist noiseless feedback communication schemes that can provide a significant improvement in terms of simplicity and reliability, see e.g. \cite{S-K_partI,S-K_partII,horstein,PM_Transactions}. However, these elegant feedback schemes completely fail in the presence of arbitrarily low feedback noise, rendering them grossly impractical. This naturally raises the question of whether simplicity and reliability can still be achieved to some degree in a practical setup of noisy feedback. In this paper, we address this question in a Gaussian setting and answer it in the affirmative. 

The setup we consider is the following. Two Terminals A and B are connected by a pair of independent AWGN channels, and are limited by individual power constraints. The channel from Terminal A (resp. B) to Terminal B (resp. A) is referred to as the feedforward (resp. feedback) channel. Terminal A is in possession of a message to be reliably transmitted to Terminal B. To that end, an interactive communication model is adopted where both terminals are allowed to employ coding and exchange signals on the fly. This model is sometimes referred to as \textit{active feedback}, and should be distinguished from the \textit{passive feedback} setting where no coding is allowed over the feedback channel. The AWGN channel with noiseless feedback was studied in the classical works of Schalkwijk and Kailath \cite{S-K_partI,S-K_partII}, who introduced a capacity-achieving communication scheme referred to herein as the \textit{S-K Scheme}. This linear-feedback coding scheme employs a first-order recursion at both terminals, and is markedly simpler than its non-feedback counterparts that typically employ long block codes and complex encoding/decoding techniques. In terms of reliability, the error probability attained by the S-K scheme decays super-exponentially with the delay, in contrast to the weaker exponential decay achieved by non-feedback codes \cite{GallagerIT}. However, this scheme and its linear feedback generalizations are not robust to any amount of feedback noise, as was initially observed in \cite{S-K_partII} and further strengthened in \cite{KimNoisyAWGNFeedbackAllertor}. 

The main contribution of this work is in showing that to some extent, the merits of noiseless feedback can be carried over to the practical regime of noisy feedback. Contrary to the noiseless feedback case, these improvements in simplicity and reliability are not simultaneously achieved. In terms of reliability, we construct two interactive protocols that are of comparable complexity to non-feedback schemes, but are superior in the asymptotic error exponent sense. In terms of simplicity, we depart from the standard asymptotic regime and show how a fixed (but low) error probability can be attained at a small capacity gap, where the latter term refers to the amount of excess $\snr$ required by the scheme above the minimum predicted by the Shannon limit. Both these constructions are useful when the signal-to-noise ratio of the feedback channel sufficiently exceeds that of the feedforward channel. 

As a case in point, consider the high-$\snr$ regime and assume that the $\snr$ of the feedback channel exceeds the $\snr$ of the feedforward channel by $20\db$. Then our simplicity-oriented scheme operates at a capacity gap of merely $0.8\db$ with only $19$ rounds of interaction, and attains a bit error rate of $10^{-6}$. This should be juxtaposed against two reference systems, operating at the same bit error rate: On the one hand, state-of-the-art non-feedback codes that attain the same capacity gap require roughly two orders of magnitude increase in delay and complexity. On the other hand, the capacity gap attained by a minimal delay uncoded system is at least $9\db$. Finally, under the same setup, our reliability-oriented schemes attain an error exponent exceeding the sphere-packing bound of the feedforward channel for a wide range of rates below capacity. 

The construction we introduce is based on endowing the S-K scheme with modulo-lattice operations. As observed in \cite{gallager2010variations}, the feedforward S-K scheme can be interpreted  as a solution to a \textit{Joint Source-Channel Coding} (JSCC) problem via analog transmission. Here, we further observe that the feedback transmission problem can be cast  as a similar problem but with side information (i.e. the message) at the receiver (i.e., Terminal A). This observation is crucial for our construction, and is leveraged by means of modulo-lattice analog transmission in the spirit of Kochman and Zamir \cite{KochmanZamirJointWZWDP}. 

Let us briefly describe the simplicity-oriented version of our scheme. Terminal A encodes its message into a scalar $\Theta$ using pulse amplitude modulation (PAM). In subsequent rounds, Terminal B computes a linear estimate of $\Theta$, and feeds back an exponentially amplified version of this estimate, modulo a fixed interval. The modulo operation facilitates the essential ``zoom-in'' amplification without exceeding the power limit, at the cost of a possible modulo-aliasing error. In turn, Terminal A employs a suitable modulo computation and obtains (if no modulo-aliasing occurs) the estimation error, corrupted by excess additive noise. This quantity is then properly scaled and sent over the feedforward channel to Terminal B. After a fixed number of rounds, Terminal B decodes the message using a minimum distance rule. Loosely speaking, the scheme's error probability is dictated by the events of a modulo-aliasing in any of the rounds, as well as the event where the remaining estimation noise is larger than the minimum distance of the PAM.  

We also introduce two asymptotic reliability-oriented versions of our scheme. The first is based on an asymptotic generalization of the simple interaction idea above, where a block code replaces the PAM modulation, and a block S-K scheme is used in conjunction with a multi-dimensional modulo-lattice operation, replacing the scalar modulo. We provide an asymptotic error exponent analysis using the Poltyrev exponent to account for modulo-aliasing errors, and channel coding exponents to account for the error of the block code. The second scheme we present is based on concatenated coding, with the scalar simplicity-oriented scheme as an inner code and a block outer code. Since the discrete memoryless channel (DMC) induced by the inner code and viewed by the outer code is non-Gaussian, we give a lower bound for the error exponent based on the performance of a ``worst case'' symmetric DMC. 

\textit{Related work.}  In \cite{KimNoisyAWGNFeedbackAllertor,BurnashevNoisyAWGNISIT}, the authors analyzed the reliability function of the AWGN channel at zero rate for noisy \textit{passive} feedback, i.e. where the channel outputs are fed back without any processing. In \cite{XiangKim}, the authors gave an interesting analysis of the reliability of transmission of an $M$-ary message ($M\geq 3$) over AWGN with passive noisy feedback, but in a slightly different setting where a peak energy constraint is imposed. In \cite{ChanceLove}, the authors considered a concatenated coding scheme with a passive linear-feedback inner code and a block outer code, and provided some error exponent results. In Section \ref{exp-discussion}, we compare our reliability-oriented scheme to \cite{ChanceLove}, and show that the exponent obtained in \cite{ChanceLove} is better for low rates whereas our exponent is better for high rates. In \cite{KimActiveFB}, which is closer to our interactive setting, the reliability function associated with the transmission of a single bit over an AWGN channel with noisy \textit{active} feedback has been considered. Specifically, it was shown that active feedback roughly quadruples the error exponent relative to passive feedback. The achievability result of \cite{KimActiveFB} is better than ours at zero rate but does not extend to positive rates.

\textit{Organization.} Notation and definitions are given in Section~\ref{sec:notat-defin}. The problem setup is introduced in Section~\ref{sec:setup}. Necessary background is given in Section~\ref{sec:perlim}. Simple interaction is addressed in Section~\ref{sec:simplicity}, and improving reliability is addressed in Section~\ref{sec:reliability}.
\nocite{GaussianInteractionISIT,SimpleInteractionAllerton2014}

\section{Notation and Definitions}\label{sec:notat-defin}
In the sequel, we use the following notation. For any number $x>0$, we write $x_{\db}\dfn10\log_{10}(x)$ to denote the value of $x$ in decibels. The Gaussian Q-function is 
\begin{align}
  \qfunc{x}\dfn\frac{1}{\sqrt{2\pi}}\int_{x}^{\infty}\exp\left( -u^2/2\right)du  
\end{align}
and $Q^{-1}(\cdot)$ is its functional inverse. We write $f(x) = \mathrm{O}(g(x))$ for  $\mathrm{limsup}_{x\to\infty} \left|f(x)/g(x)\right| < \infty$. We write $\log$ for base $2$ logarithm, and $\ln$ for the natural logarithm. We use the vector notation $\bs{x}^n\dfn (x_1,\ldots,x_n)$ and boldface letters such as $\bM{x}$ to indicate vectors of size $\latticeN$. We write $a_n\dotgeq b_n$ to mean  $\liminf_{n\rightarrow\infty}\frac{1}{n}\ln\left(\frac{a_n}{b_n}\right)\geq 0$, and similarly define $\dotleq$ and $\doteq$. We use $\overline{A}$ to denote the complementary of an event $A$.

\textit{The Capacity Gap}. Recall that the Shannon capacity of the AWGN channel with signal-to-noise ratio $\snr$ is given by
\begin{align}
\label{eq:AWNGcap}
C = \frac{1}{2}\log(1+\snr).
\end{align}
This is the maximal rate achievable by any scheme (of unbounded complexity/delay, with or without feedback) under vanishing error probability. Conversely, the minimal $\snr$ required to reliably attain a rate $R$ is $2^{2R}-1$. The \textit{capacity gap} $\Gamma$ attained by a coding scheme that operates at rate $R$ over an AWGN channel, is the  excess $\snr$ required by the scheme over the minimum predicted by the Shannon limit, i.e., 
\begin{align}
\label{eq:capGapDef}
  \Gamma \dfn \frac{\snr}{2^{2R}-1}.
\end{align}
Note that if a nonzero bit/symbol error probability is allowed, then one can achieve rates exceeding the Shannon capacity \eqref{eq:AWNGcap}, and this effect should in principle be accounted for, to make the definition of the capacity gap fair. However, for small error probabilities the associated correction factor (related to the corresponding rate-distortion function) becomes negligible, and we therefore ignore it in the sequel. 

\section{Setup}\label{sec:setup}
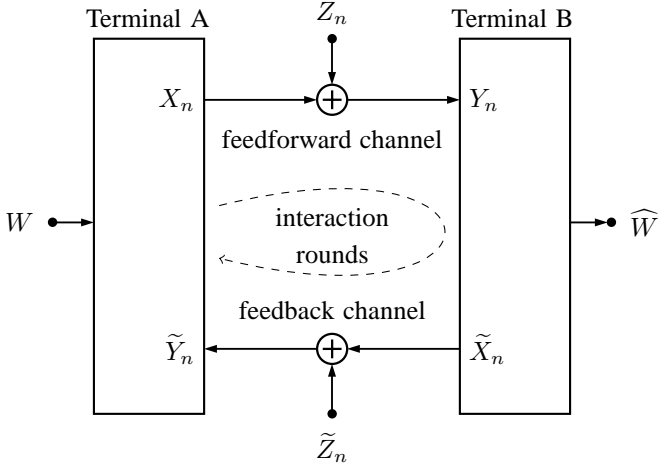
\begin{figure}
\centering
\begin{tikzpicture}
[every text node part/.style={align=center}]

	\matrix (m1) [row sep=2.5mm, column sep=5mm]
	{  

                \node[coordinate] (){};&
                \node[coordinate] (){};&
                \node[coordinate] (){};&
                \node[coordinate] (r1ur){};&
                \node[coordinate] (){};&
                \node[coordinate] (){};&
                \node[dspnodefull,dsp/label=above]   (z1){${Z}_n$};&
                \node[coordinate] (){};&
                \node[coordinate] (){};&
                \node[coordinate] (r2ur){};&
                \\
                \\
                \node[coordinate] (){};&
                \node[coordinate] (){};&
                \node[coordinate] (){};&
		\node[left]                  (x1) {$X_n$};   &
                \node[coordinate] (){};&
                \node[coordinate] (){};&
		\node[dspadder,dsp/label=below]  (a1) {feedforward channel}; &
                \node[coordinate] (){};&
                \node[coordinate] (){};&
		\node[right]                  (y1) {$Y_n$};  &

		\\ \\ \\ \\ \\
                \node[coordinate] (){};&
                \node[dspnodefull,dsp/label=left]   (win){$W$};&
                \node[right] (wout){};&
                \node[coordinate] (){};&
                \node[coordinate] (){};&
                \node[coordinate] (){};&
                \node[coordinate] (){};&
                \node[coordinate] (){};&
                \node[coordinate] (){};&
                \node[coordinate] (){};&
                \node[left] (whin){};&
                \node[dspnodefull,dsp/label=right]   (whout){$\widehat{W}$};&
                \\ \\ \\ \\ \\

                \node[coordinate] (){};&
                \node[coordinate] (){};&
                \node[coordinate] (){};&
                \node[left] (y2){$\widetilde{Y}_n$};&
                \node[coordinate] (){};&
                \node[coordinate] (){};&
                \node[dspadder,dsp/label=above]   (a2){feedback channel};&
                \node[coordinate] (){};&
                \node[coordinate] (){};&
                \node[right] (x2){$\widetilde{X}_n$};&
                \\  \\

                \node[coordinate] (){};&
                \node[coordinate] (){};&
                \node[coordinate] (r1ld){};&
                \node[coordinate] (){};&
                \node[coordinate] (){};&
                \node[coordinate] (){};&
                \node[dspnodefull,dsp/label=below]   (z2){$\widetilde{Z}_n$};&
                \node[coordinate] (){};&
                \node[coordinate] (){};&
                \node[coordinate] (){};&
                \node[coordinate] (r2ld){};&
                \node[coordinate] (){};&\\ \\
	};
        
          \draw[dspconn] (x1) -- (a1);
          \draw[dspconn] (a1) -- (y1);
          \draw[dspconn] (z1) -- (a1);
          \draw[dspconn] (x2) -- (a2);
          \draw[dspconn] (a2) -- (y2);
          \draw[dspconn] (z2) -- (a2);
          \draw[dspconn] (win) -- (wout);
          \draw[dspconn] (whin) -- (whout);

	\draw [color=black,thick](r1ur) rectangle (r1ld);
	\node at (-2.45,2.9) [] {\textrm{Terminal A}};

	\draw [color=black,thick](r2ur) rectangle (r2ld);
	\node at (2.4,2.9) [] {\textrm{Terminal B}};

       \def\xoffset{-10}
       \def\yoffset{1.2}
        \draw [->,dashed] (8.5+\xoffset,-0.8+\yoffset) .. controls (12.5+\xoffset,0.2+\yoffset) and (12.5+\xoffset,-2.5+\yoffset) .. (8.5+\xoffset,-1.5+\yoffset);
        \node at (0,0)  {{interaction} \\ {rounds}};

\end{tikzpicture}
\caption{\label{fig:blockdiagram}Block diagram of interactive coding over an AWGN channel with noisy feedback}
\end{figure}

Our problem setup is depicted in Fig.~\ref{fig:blockdiagram}. The feedforward and feedback channels connecting Terminal A to Terminal B and vice versa, are AWGN channels given by 
\begin{align}
Y_n=X_n+Z_n ,  \\
\wt{Y}_n=\wt{X}_n+\wt{Z}_n,
\end{align}
where $X_n, Y_n$ (resp. $\wt{X}_n,\wt{Y}_n$) are the input and output of the feedforward (resp. feedback) channel at time $n$ respectively. The feedforward (resp. feedback) channel noise $Z_n\sim \mathcal{N}(0,\sigma^2)$ (resp. $\wt{Z}_n\sim \mathcal{N}(0,\wt{\sigma}^2)$) is independent of the input $X_n$ (resp. $\wt{X}_n$), and constitutes an i.i.d. sequence. The feedforward and feedback noise processes are mutually independent. 

Terminal A is in possession of a message $W$, uniformly distributed over the set $\{0,...,M-1\}$, to be described to Terminal B over $N$ rounds of communication. To that end, the terminals can employ an interactive scheme defined by the sequences of functions $\varphi_n$ and $\wt{\varphi}_n$ as follows: At time $n$, Terminal A sends a function of its message $W$ and possibly of past feedback channel outputs over the feedforward channel, i.e., 
\begin{align}
  X_n=\varphi_n(W,\wt{Y}^{n-1}).
\end{align}
Similarly, Terminal B sends function of its past observations to Terminal A over the feedback channel, i.e., 
\begin{align}
  \wt{X}_n=\wt{\varphi}_n(Y^n).
\end{align}
\begin{remark}
In general, we allow these functions to further depend on common randomness shared by the terminals. We note in passing that our definition of the feedback transmission scheme is sometimes referred to as \textit{active feedback}; the term \textit{passive feedback} is usually reserved to the special case where $\wt{\varphi}(Y^n)=Y_n$. 
\end{remark}

The number of rounds $N$ is fixed. While feedback protocols with variable transmission length exist and can improve reliability relative to non-feedback transmission \cite{Burnashev,SatoYamamoto}, they are beyond the scope of this work. We assume that Terminal A (resp. Terminal B) is subject to an average power constraint $P$ (resp. $\wt{P}$), namely
\begin{align}
\sum_{n=1}^N\mathbb{E}(X_n^2) \leq N\cdot P, \quad \sum_{n=1}^N\mathbb{E}(\wt{X}_n^2) \leq N\cdot \wt{P} .
\end{align}
We denote the feedforward (resp. feedback) signal-to-noise ratio by $\snr\dfn\frac{P}{\sigma^2}$ 
(resp.  $\bsnr\dfn \frac{\wt{P}}{\wt{\sigma}^2}$). The excess signal-to-noise ratio of the feedback over the feedforward is denoted by $\dsnr\dfn\frac{\bsnr}{\snr}$. Throughout this work, we assume that $\dsnr > 1$.

An interactive scheme $(\varphi,\wt{\varphi})$ is associated with a rate $R\dfn \frac{\log{M}}{N}$ and an error probability $\Pe$, which is the probability that Terminal B errs in decoding the message $W$ at time $N$, under the optimal decision rule.

\section{Preliminaries}\label{sec:perlim}
In this section, we describe the building blocks underlying our interactive scheme. First, we review the use of uncoded PAM signaling, and discuss its associated capacity gap. Then, we describe the basic problem of joint source-channel coding (JSCC) via analog transmission, and show how to build the S-K scheme from uncoded PAM and iterative JSCC. Lastly, we discuss the problem of JSCC with side information using modulo arithmetic, and present a simple scalar solution that is later implemented as part of our simplicity-oriented scheme. 

\subsection{Uncoded PAM} 
\label{pambasic}
PAM is a simple and commonly used modulation scheme, where $2^R$ symbols are mapped (one-to-one) to the set 
\begin{align}
\{\pm 1\eta,\pm 3\eta,\cdots,\pm (2^R-1)\eta\}.
\end{align}
Canonically, the normalization factor $\eta$ is set so that the overall mean square of the constellation (assuming equiprobable symbols) is unity. A straightforward calculation yields $\eta = \sqrt{3/\left( 2^{2R}-1\right)}$. In the general case where the mean square of the constellation is constrained to be $P$, $\eta$ is replaced with $\eta\sqrt{P}$. 

It is easy to show that for an AWGN channel with zero mean noise of variance $\sigma^2$ and average input power constraint $P$, the probability of error incurred by the optimal detector is given by the following formula
\begin{align}
\Pe&=2\left(1-2^{-R}\right)\qfunc{\frac{\sqrt{P}\eta}{\sigma}}\\ 
\label{eq:PeUncodedPAM}
&=2\left(1-2^{-R}\right)\qfunc{\sqrt{\frac{3\snr}{2^{2R}-1}}}.
 \end{align}
Manipulating \eqref{eq:PeUncodedPAM} yields:
\begin{align}
\label{eq:capGapaccurate}
R=\frac{1}{2}\log\left(1+\frac{\snr}{\frac{1}{3}\left[Q^{-1}\left(\frac{\Pe}{\left(1-2^{-R}\right)}\right)\right]^2} \right),
\end{align}
which can be slightly relaxed to obtain a lower bound on $R$ by
\begin{align}
\label{eq:capGap}
R>\frac{1}{2}\log\left(1+\frac{\snr}{\Gamma_0(\Pe)} \right),
\end{align}
where
\begin{align}
\label{eq:gammaPAM}
\Gamma_0(\Pe)\dfn\frac{1}{3}\left[Q^{-1}\left(\frac{\Pe}{2}\right)\right]^2.  
\end{align}
Comparing \eqref{eq:capGap} and \eqref{eq:AWNGcap}, we see that PAM signaling with error probability $\Pe$ admits a capacity gap of at most $\Gamma_0(\Pe)$. Comparing \eqref{eq:capGapaccurate} and \eqref{eq:capGap} it is clear that this upper bound is tight as $R$ increases.
For a typical value of $\Pe=10^{-6}$, the capacity gap of uncoded PAM is at most $\Gamma_{0,\db} = 9\db$. 

Finally, we assume as usual that bits are mapped to PAM constellation symbols via  Gray labeling. The associated bit error probability can thus be bounded by
\begin{align}\label{eq:pb}
p_b<\frac{2}{R}\qfunc{\frac{\sqrt{P}\eta}{\sigma}}+2\qfunc{3\frac{\sqrt{P}\eta}{\sigma}} \approx \frac{\Pe}{R}.
\end{align}
The bound follows by noting that erring toward the nearest neighbor incurs an error in a single bit, 
and by taking a worst case assumption for all other error events. The approximation becomes tight for small $\Pe$ due to the strong decay of the Q-function. 

\subsection{Joint Source-Channel-Coding (JSCC) via Analog Transmission}
It is well known \cite{Goblick} that when a Gaussian source is to be transmitted over an AWGN channel under a quadratic distortion measure, analog transmission obtains the optimal distortion (given by equating the rate-distortion function to the channel capacity) with minimal delay. This solution is a simple instance of joint source-channel coding (JSCC). More explicitly, we wish to convey a Gaussian r.v. $\eps\sim\mathcal{N}\left(0,\sigma^2_{\eps}\right)$ over an AWGN channel $Y=X+Z$, $Z\sim \mathcal{N}\left(0,\sigma^2 \right)$, with expected input power constraint $\Expt{X^2}\leq P$ (i.e. $\snr=\frac{P}{\sigma^2}$). The optimal transmission and estimation boil
 down to $X=\alpha \eps$ and $\wh{\eps}=\beta Y$. The optimal choice of $\alpha$ yields a power scaling factor, i.e. $\alpha=\frac{\sqrt{P}}{\sigma_{\eps}}$, and optimal choice of $\beta$ yields the Wiener coefficient $\beta=\frac{\sigma_{\eps}}{\sigma}\frac{\sqrt{\snr}}{\snr+1}$. Plugging $\alpha$ and $\beta$ yields the minimal attainable MSE  in this setup:
\begin{align}
\label{eq:analogTrans}
\Expt{\left(\wh{\eps}-\eps\right)^2}=\frac{\sigma_{\eps}^2}{\snr+1}
\end{align}
Namely, this JSCC scheme improves the estimation error of $\eps$ by a factor $\snr+1$ relative to a trivial guess. In the sequel we shall use this simple construction as a building block for both the classic S-K scheme and the newly proposed noisy feedback schemes.

\subsection{The S-K Scheme}
\label{s-kbasic}
Consider the setting of communication over the AWGN channel with noiseless feedback, i.e., where $\wt{\sigma}^2 = 0$. The S-K scheme can be described as follows. First, Terminal A  maps the message $W$ to the real-valued variable $\Theta$ using a PAM modulation of size $2^{NR}$. In the first round, it sends a scaled version of $\Theta$ satisfying the power constraint $P$. In subsequent rounds, Terminal B maintains an estimate $\wh{\Theta}_n$ of $\Theta$ given all the observation it has, and feeds it back to Terminal A. Terminal A then computes the estimation error $\eps_n\dfn \wh{\Theta}_n-\Theta$, and sends it to Terminal B using analog transmission.
\begin{enumerate}[(A)]
\item Initialization:
\begin{enumerate}[]
\item \textbf{Terminal A:} Map the message $W$ to a PAM point $\Theta$.  
\item \textbf{Terminal A $\Rightarrow$ Terminal B:} 
  \begin{itemize}
  \item Send $X_1=\sqrt{P}\Theta$
    \item Receive $Y_1=X_1+Z_1$
  \end{itemize}  

\item \textbf{Terminal B:} Initialize the $\Theta$ estimate\footnote{\label{fn1}Note that this is the minimum variance unbiased estimate of $\Theta$.}
to $\wh{\Theta}_1=\frac{Y_1}{\sqrt{P}}$. 
\end{enumerate}
\item Iteration:
\begin{enumerate}[]
\item \textbf{Terminal B $\Rightarrow$ Terminal A:} 
  \begin{itemize}
  \item Send the current $\Theta$ estimate: $\wt{X}_n= \wh{\Theta}_n$
  \item Receive $\wt{Y}_n=\wt{X}_n$
  \end{itemize}  
\item \textbf{Terminal A:} Compute the estimation error $\varepsilon_n = \wt{Y}_n - \Theta$.
\item \textbf{Terminal A $\Rightarrow$ Terminal B:} 
  \begin{itemize}
  \item Send $\eps_n$ via analog transmission. i.e. $X_{n+1}=\alpha_n \varepsilon_n$, where $\alpha_n=\frac{\sqrt{P}}{\sigma_n}$ where $\sigma_n^2\dfn \Expt{\eps_n^2}$.
  \item Receive $Y_{n+1}=X_{n+1}+Z_{n+1}$
  \end{itemize}  
\item \textbf{Terminal B:} 
Update the $\Theta$ estimate\footref{fn1} $\wh{\Theta}_{n+1}=\wh{\Theta}_n-\wh{\varepsilon}_n$, where  
\begin{align}\label{eq:eps_est}
\wh{\varepsilon}_n=\beta_{n+1}Y_{n+1}  
\end{align}
is the \textit{Minimum Mean Squared Error} (MMSE) estimate of $\varepsilon_n$, thus, $\beta_{n+1}$ is the appropriate Wiener coefficient:
\begin{align}
\beta_{n+1}=\frac{\sqrt{P\sigma_n^2}}{P+\sigma^2}=\frac{\sigma_n}{\sigma}\cdot\frac{\sqrt{\snr}}{1+\snr}.
\end{align}
\end{enumerate}
\item Decoding: 

At time $N$, Terminal B decodes the message using a minimum distance decoder for $\wh{\Theta}_N$ w.r.t. the PAM constellation. 
\end{enumerate}

To calculate the error probability and rate attained by the S-K scheme, we note that $\eps_{n+1} = \eps_n - \wh{\eps}_n$. Using the property \eqref{eq:analogTrans} of analog transmission yields:
\begin{align}
\label{eq:sigman}
\sigma_{n+1}^2=\frac{\sigma_{n}^2}{1+\snr}=\frac{1}{\snr\left(1+\snr\right)^{n}}.
\end{align}
An important observation is that using \eqref{eq:sigman} and the fact that the power of $\Theta$ is normalized to unity, one can regard the channel from $\Theta$ to $\Theta_N$ as an AWGN channel with a signal-to-noise ratio ${\snr}_N=\sigma_N^{-2}$, namely:
\begin{align}
\label{eq:snreq}
\snr_N=\snr\cdot(1+\snr)^{N-1}.
\end{align}
Note that it is possible to use the (biased) MMSE at the first round and compensate for the bias in the last round \cite{ElGamalKimBook}, yielding an end-to-end signal-to-noise ratio of $(1+\snr)^{N}-1$. However, this improvement is negligible and also complicates the analysis in the sequel.
Plugging $\snr_N$ into \eqref{eq:PeUncodedPAM} and bounding the Q-function by 
$Q(x)<\frac{1}{2}\exp(-\frac{1}{2}x^2)$ gives:
\begin{align}
\Pe<\exp\left(-\frac{3}{2}\frac{\snr\cdot(1+\snr)^{N-1}}{2^{2NR}-1}\right).
\end{align}
Plugging in the AWGN channel capacity \eqref{eq:AWNGcap} and removing the ``$-1$'' term, we obtain:
\begin{align}
\Pe<\exp\left(-\tfrac{3}{2}\tfrac{\snr}{1+\snr} \cdot 2^{2N(C-R)}\right).
\end{align}
which is the well-known doubly exponential decay of the error probability of the S-K scheme. 

Let us now provide an alternative interpretation of the S-K scheme performance, in terms of the capacity gap attained after a finite number of rounds. Plugging $\snr_N$ in \eqref{eq:capGap} yields:
\begin{align}
\label{eq:targetRate}
R>\frac{1}{2N}\log\left(1+\frac{\snr\cdot(1+\snr)^{N-1}}{\Gamma} \right).
\end{align}
Substituting the resulting $R$ in the definition of the capacity gap \eqref{eq:capGapDef} 
and assuming $\snr\gg 1$ yields the following approximation for high $\snr$:
\begin{align}
  \Gamma_{\db}^{\text{S-K}}(\Pe,N) \approx \frac{\Gamma_{0,\db}(\Pe)}{N}.
\end{align}
This behavior is depicted by the dashed curve in Fig.~\ref{fig:resultsFigR4}.

\subsection{Joint Source-Channel Coding with Side Information  \label{sec:joint-source-channel}}
A key observation made in this work is that while the transmission of $\eps_n$ over the feedforward link (i.e. from Terminal A to Terminal B) can be regarded as a JSCC problem, the transmission of  $\wh{\Theta}_n$ over the feedback link (i.e. from Terminal B to Terminal A) can be regarded as a JSCC problem with side information. More explicitly, at round $n$, Terminal B holds its current estimate $\wh{\Theta}_n=\Theta+\eps_n$ and wants to convey it to Terminal A, while Terminal A knows $\Theta$ and can use it as side information. To exploit this, we employ a lattice-based JSCC scheme with side information based on a more general scheme by Kochman and Zamir \cite{KochmanZamirJointWZWDP}. We note that for clarity of exposition and ease of analysis, we use the high-$\snr$ version of \cite{KochmanZamirJointWZWDP}, which can be slightly suboptimal in the low-$\snr$ regime. 

Let us now describe this solution. For simplicity, we start with the scalar case. First, we need some definitions and properties of modulo arithmetic. For a given  $d>0$, the scalar modulo-$d$ function is 
\begin{align}
\moduloOp{x}\dfn x-d\cdot\textrm{round}\left(\frac{x}{d}\right)  
\end{align}
where the \textrm{round$(\cdot)$} operator returns nearest integer to its argument (rounding up at half). The following properties are easily verified. 
\begin{proposition}\label{prop:scalarMod} 
The following properties hold:	
  \begin{enumerate}[(i)]
  \item $\moduloOp{x}\in[-\frac{d}{2},\frac{d}{2})$
  \item \label{ii} The \textit{modulo distributive law} 
	  \begin{align}
	  \moduloOp{\moduloOp{x+d_1}+d_2-x}=\moduloOp{d_1+d_2}
	  \end{align}
  \item \label{iii} if $d_1+d_2\in[-\frac{d}{2},\frac{d}{2})$, then 
    \begin{align}\label{eq:modrel}
      \moduloOp{\moduloOp{x+d_1}+d_2-x}=d_1+d_2.
    \end{align}
    otherwise, a \textit{modulo-aliasing error} term of $k d$ is added to the right-hand-side \eqref{eq:modrel}, for some integer $k\neq0$. 
  \item Let $V\sim\textrm{Uniform}([-\frac{d}{2},\frac{d}{2}))$. Then $\moduloOp{x+V}$ is uniformly distributed over $[-\frac{d}{2},\frac{d}{2})$ for any $x\in\RealF$.
  \item $\mathbb{E}(\moduloOp{x+V})^2=\frac{d^2}{12}$. 
  \end{enumerate}
\end{proposition}

Using the above properties, we can provide the following solution to the JSCC with side information problem over the feedback link. Terminal B is in possession of $\wh{\Theta}=\Theta+\eps$ and wishes to convey it to Terminal A, where  $\eps\sim\mathcal{N}(0,\sigma^2_\eps)$ and $\Theta$ is known to Terminal A. The terminals are connected through an AWGN channel: $\wt{Y}=\wt{X}+\wt{Z}$, with the appropriate noise distribution $\wt{Z}\sim \mathcal{N}(0,\wt{\sigma}^2)$ and an input power constraint $\Expt{\wt{X}^2}\leq \wt{P}$. Let $V\sim\textrm{Uniform}([-\frac{d}{2},\frac{d}{2}))$  be a \textit{dither} signal known at both terminals. Then, Terminal B transmits 
\begin{align}
  \wt{X} = \moduloOp{\gamma\wh{\Theta}+V},
\end{align}
where we set $d=\sqrt{12\wt{P}}$ to guarantee that the power constraint is satisfied. Terminal A computes the estimate
\begin{align}
  \wh{\eps}=\frac{1}{\gamma}\moduloOp{\wt{Y}-\gamma\Theta-V}.
\end{align}
Hence, by Proposition~\ref{prop:scalarMod} property (\ref{ii}):
\begin{align}\label{eq:moduloDec}
  \wh{\eps}=\frac{1}{\gamma}\moduloOp{\gamma\eps+\wt{Z}}.
\end{align}
In the case where $\gamma\eps+\wt{Z}\in[-\frac{d}{2},\frac{d}{2})$ (Proposition~\ref{prop:scalarMod} property (\ref{iii})) we obtain
\begin{align}\label{eq:epshat}
  \wh{\eps}=\eps+\frac{1}{\gamma}\wt{Z}.
\end{align}
The question that arises at this point is how to set $\gamma$. Clearly, a large $\gamma$ would increase the probability of a modulo-aliasing error, but at the same time would reduce the additive estimation error in $\wh{\eps}$. Denoting the modulo-aliasing error probability by $\Pmod$. Proposition~\ref{prop:scalarMod} property (\ref{iii}) implies that 
\begin{align}
  \Pmod\dfn\Pr\left(\gamma\eps+\wt{Z}\notin[-\tfrac{d}{2},\tfrac{d}{2})\right).
\end{align}
Recalling that $d=\sqrt{12\wt{P}}$, and that $\eps$ and $\wt{Z}$ are jointly Gaussian with known variances, we obtain
\begin{align}\label{eq:pmodscalar}
  \Pmod=2Q\left(\sqrt{\frac{3\wt{P}}{\gamma^2\sigma^2_{\eps}+\wt{\sigma}^2}}{}\right).
\end{align}
Let us now introduce the \textit{looseness parameter} $\latticeLoose$, defined as:
\begin{align}
\label{eq:latticecLooseDef}
  \latticeLoose = \frac{\wt{P}}{\gamma^2\sigma^2_\eps+\wt{\sigma}^2}. 
\end{align}
Using this definition, we can write $\Pmod$ as 
\begin{align}
\Pmod=2Q\left(\sqrt{3\latticeLoose}\right)
\end{align}
Observe that a larger $\latticeLoose$ implies a smaller variance of the modulo argument in \eqref{eq:moduloDec}, hence  a smaller modulo-aliasing error probability. On the other hand, a larger $\latticeLoose$ implies a smaller $\gamma$, and hence a larger estimation error by virtue of~\eqref{eq:epshat}. In the sequel, it will be convenient to express our results in terms of $L$ instead of $\gamma$, as the former is a more natural parameter of the problem. 

\section{Simple Interaction\label{sec:simplicity}}
In this section we present our simplicity-oriented interactive scheme, using the S-K scheme and the scalar modulo JSCC scheme with side information as building blocks. We analyze the associated capacity gap and discuss implementation issues. The scheme is presented in Subsection~\ref{ourscheme}. An upper bound on the capacity gap attained by the scheme is given in Subsection~\ref{sec:main-res}, and proved in Subsection~\ref{sec:proof}. Numerical results are presented in Subsection~\ref{results}. Practical implementation issues are addressed in Subsection~\ref{sec:implem}, and a concluding discussion appears in Subsection~\ref{sec:discussion}. 

\subsection{The Proposed Scheme}\label{ourscheme}
\begin{figure*}
\centering
\begin{tikzpicture}

	\matrix (m1) [row sep=2.5mm, column sep=5mm]
	{
                \node[coordinate] (){};&
                \node[coordinate] (){};&
                \node[dspnodefull,above] (mf1){$-V_{n}$};&
                \node[coordinate] (){};&
                \node[coordinate] (){};&
                \node[coordinate] (){};&
                \node[coordinate] (){};&
                \node[dspnodefull,dsp/label=above]   (mf4){${Z}_n$};&
                \node[coordinate] (){};&
                \node[coordinate] (){};&
                \node[coordinate] (){};&
                \node[coordinate] (mf6){};&
                \node[dspsquare] (mf7){$D$};&
		\node[coordinate] (mf8){}; &
                \node[coordinate] (){};&\\

		\node[dspnodeopen,dsp/label=left] (m0f) {$\Theta$};    &
		\node[dspmixer,dsp/label=above]   (m00) {$-\gamma_{n}$};          &
		\node[dspadder]                    (m01) {};          &
		\node[dspsquare, inner xsep=1pt]                   (m02) {$\mathbb{M}_d(\cdot)$}; &
		\node[dspmixer]                    (m03) {$\alpha$};    &
		\node[coordinate]                  () {};          &
		\node[above]                  () {$X_n$};          &
		\node[dspadder,dsp/label=below]  (m04) {{feedforward channel}}; &
		\node[above]                  () {$Y_n$};          &
		\node[coordinate]                  () {};          &
		\node[dspmixer]                    (m05) {$-\beta_n$};          &
		\node[dspadder]                    (m06) {}; &
		\node[coordinate]                  (m07) {};          &
		\node[dspnodefull] (m08) {};          &
		\node[right] (m09) {$\widehat{\Theta}_n$};          &
		\\
                \node[coordinate] (){};&
                \node[coordinate] (){};&
                \node[coordinate] (m11){};&
                \node[coordinate] (){};&
                \node[coordinate] (){};&
                \node[coordinate] (){};&
                \node[coordinate] (){};&
                \node[coordinate] (){};&
                \node[coordinate] (){};&
                \node[coordinate] (){};&
                \node[coordinate] (){};&
                \node[coordinate] (){};&
                \node[coordinate] (){};&
                \node[dspmixer,dsp/label=left]  (m18){$\gamma_n$};&
                \node[coordinate] (){};&\\

                \node[coordinate] (){};&
                \node[coordinate] (){};&
                \node[coordinate] (){};&
                \node[coordinate] (){};&
                \node[coordinate] (){};&
                \node[coordinate] (){};&
                \node[coordinate] (){};&
                \node[coordinate] (){};&
                \node[coordinate] (){};&
                \node[coordinate] (){};&
                \node[coordinate] (){};&
                \node[coordinate] (){};&
                \node[dspnodefull,dsp/label=left] (md7){$V_n$};&
                \node[dspadder]   (md8){};&
                \node[coordinate] (){};&
                \\

		--------------------------------------------------------------------
                \node[coordinate] (){};&
                \node[coordinate] (){};&
                \node[coordinate] (){};&
                \node[coordinate] (){};&
                \node[coordinate] (){};&
                \node[coordinate] (){};&
                \node[coordinate] (){};&
                \node[coordinate] (){};&
                \node[coordinate] (){};&
                \node[coordinate] (){};&
                \node[coordinate] (){};&
                \node[coordinate] (){};&
                \node[coordinate] (){};&
		\node[dspsquare, inner xsep=1pt]  (m28){$\mathbb{M}_d(\cdot)$}; &
                \node[coordinate] (){};&\\
                \node[coordinate] (){};&
                \node[coordinate] (){};&
                \node[coordinate] (m31){};&
                \node[coordinate] (){};&
                \node[coordinate] (){};&
                \node[coordinate] (){};&
                \node[below] (){$\widetilde{Y}_n$};&
                \node[dspadder,dsp/label=above]   (m34){{feedback channel}};&
                \node[below] (){$\widetilde{X}_n$};&
                \node[coordinate] (){};&
                \node[coordinate] (){};&
                \node[coordinate] (){};&
                \node[coordinate] (){};&
		\node[coordinate] (m38){}; &
                \node[coordinate] (){};&
                \\
                \\
                \node[coordinate] (){};&
                \node[coordinate] (){};&
                \node[coordinate] (){};&
                \node[coordinate] (){};&
                \node[coordinate] (){};&
                \node[coordinate] (){};&
                \node[coordinate] (){};&
                \node[dspnodefull,dsp/label=below]   (m44){$\widetilde{Z}_n$};&
                \node[coordinate] (){};&
                \node[coordinate] (){};&
                \node[coordinate] (){};&
                \node[coordinate] (){};&
                \node[coordinate] (){};&
		\node[coordinate] (){}; &
                \node[coordinate] (){};&\\ \\
	};
        
	\begin{scope}[start chain]
	\chainin (m00);
         \foreach \i in {1,2,3,4,5,6,9}
         {
		\chainin (m0\i) [join=by dspconn];
         }
        \chainin (m08);
        \chainin (m18) [join=by dspconn];
        \chainin (md8) [join=by dspconn];
        \chainin (m28) [join=by dspconn];
        \chainin (m38) [join=by dspline];
        \chainin (m34) [join=by dspconn];
        \chainin (m31) [join=by dspline];
        \chainin (m01) [join=by dspconn];
        \draw[dspconn] (m44) -- (m34);
        \draw[dspconn] (mf4) -- (m04);
        \draw[dspline] (m18) -- (mf8);
        \draw[dspconn] (mf8) -- (mf7);
        \draw[dspline] (mf7) -- (mf6);
        \draw[dspconn] (mf6) -- (m06);
        \draw[dspflow] (m06) -- (m08);
        \draw[dspconn] (mf1) -- (m01);
        \draw[dspconn] (m0f) -- (m00);
        \draw[dspconn] (md7) -- (md8);

	\end{scope}

	\draw [color=gray,thick](-7.1,-2) rectangle (-2.7,3.4);
	\node at (-6.3,3.7) [] {\textrm{Terminal A}};

	\draw [color=gray,thick](1.3,-2) rectangle (5.85,3.4);
	\node at (5,3.7) [] {\textrm{Terminal B}};

\end{tikzpicture}
\caption{\label{fig:blockdiagramExplicit} Block diagram of the proposed scheme}
\end{figure*}
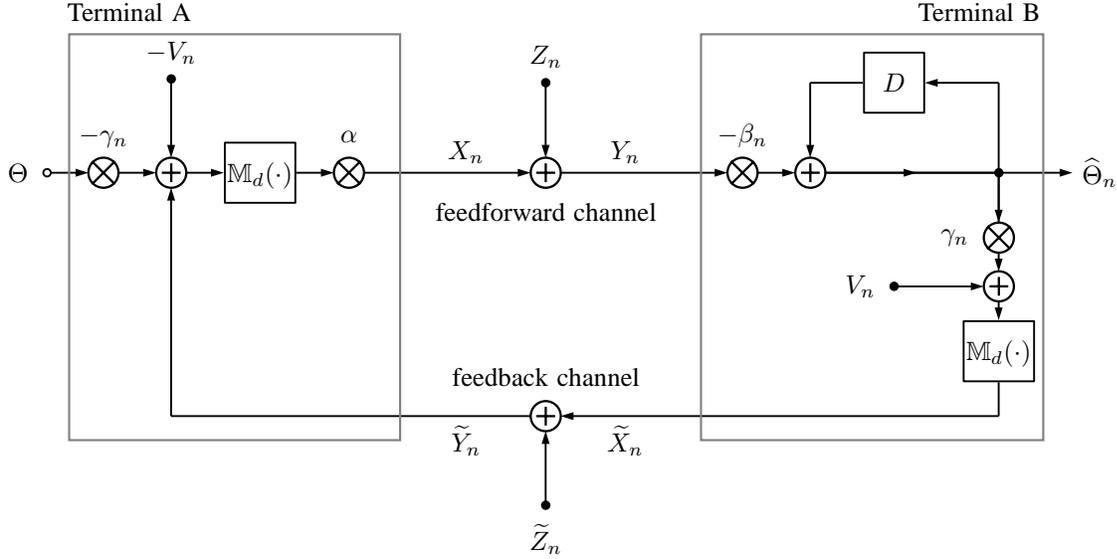

In what follows, we assume that the terminals share a common random i.i.d sequence $\{V_n\}_{n=1}^N$, mutually independent of the noise sequences and the message, where $V_n\sim\textrm{Uniform}([-\frac{d}{2},\frac{d}{2}))$. As before, we set $d=\sqrt{12\wt{P}}$. Recall the definition of $\wh{\Theta}_n$ and $\eps_n$ in Subsection~\ref{s-kbasic}, as the estimator of $\Theta$ and the corresponding estimation error at Terminal B at time $n$. 

A block diagram of the scheme is depicted in Fig.~\ref{fig:blockdiagramExplicit}. Let us describe our scheme in detail. The scheme is given in terms of the parameters $\alpha,\beta_n,\gamma_n$ which dictate the performance. The specific choice of these parameters is given in the next subsection. 
\begin{enumerate}[(A)]
\item Initialization:
\begin{enumerate}[]
\item \textbf{Terminal A:} Map the message $W$ to a PAM point $\Theta$.  
\item \textbf{Terminal A $\Rightarrow$ Terminal B:} 
  \begin{itemize}
  \item Send $X_1=\sqrt{P}\Theta$
    \item Receive $Y_1=X_1+Z_1$
  \end{itemize}  

\item \textbf{Terminal B:} Initialize the $\Theta$ estimate to $\wh{\Theta}_1=\frac{Y_1}{\sqrt{P}}$. 
\end{enumerate}
\item Iteration:
\begin{enumerate}[]
\item \textbf{Terminal B $\Rightarrow$ Terminal A:} 
  \begin{itemize}
  \item Given the $\Theta$ estimate $\wh{\Theta}_n$, compute and send 
    \begin{align}
      \wt{X}_n= \moduloOp{\gamma_n\wh{\Theta}_n+V_n}
    \end{align}
  \item Receive $\wt{Y}_n=\wt{X}_n + \wt{Z}_n$
  \end{itemize}  
\item \textbf{Terminal A:} Extract a noisy version of
estimation error $\varepsilon_n$:
\begin{align}\label{eq:noisy_estimate}
\wt{\varepsilon}_n&=\frac{1}{\gamma_n}\moduloOp{\wt{Y}_n-\gamma_n{\Theta}-V_n}\\
\label{eq:epsntilde}
&=\frac{1}{\gamma_n}\moduloOp{ \gamma_n\eps_n+\wt{Z}_n} 
\end{align}
Note that $\wt{\varepsilon}_n = \eps_n + \frac{1}{\gamma_n}\wt{Z}_n$, unless a \textit{modulo-aliasing error} occurs. 
 \item \textbf{Terminal A $\Rightarrow$ Terminal B:} 
  \begin{itemize}
  \item Send a scaled version of $\wt{\varepsilon}_n$:
$X_{n+1}=\alpha\gamma_n\wt{\varepsilon}_n$, where $\alpha$ is set to satisfy input power constraint $P$ (computed later). 
  \item Receive $Y_{n+1}=X_{n+1}+Z_{n+1}$
  \end{itemize}  
\item \textbf{Terminal B:} 
Update the $\Theta$ estimate $\wh{\Theta}_{n+1}=\wh{\Theta}_n-\wh{\varepsilon}_n$, where  
\begin{align}
\wh{\varepsilon}_n=\beta_{n+1}Y_{n+1}  
\end{align}
The choice of $\beta_n$ is described in the sequel. 
\end{enumerate}
\item Decoding: 

At time $N$ the receiver decodes the message using a minimum distance decoder for $\wh{\Theta}_N$ w.r.t. the PAM constellation. 
\end{enumerate}

\subsection{Main Result: The Capacity Gap}\label{sec:main-res}
Recall the capacity gap function $\Gamma_0(\cdot)$ of uncoded PAM given in \eqref{eq:gammaPAM}. Fix a desired number of rounds $N$, a target error probability $\Pet$, and set the target modulo-error probability to 
\begin{align}
\label{eq:pem}
p'_m=\frac{\Pet}{2(N-1)}.
\end{align}
Set the looseness parameter to 
\begin{align}\label{eq:Ldef}
  \latticeLoose &= \frac{1}{3}\left[Q^{-1}\left(\frac{p'_m}{2}\right)\right]^{2}, 
\end{align}
and set the scheme parameters $\alpha,\beta_n,\gamma_n$ to
\begin{align}
\alpha=\sqrt{\psid^{-1}\latticeLoose\frac{P}{\wt{P}}},
\end{align}
\begin{align} 
\beta_{n}=\frac{\sigma_{n-1}}{\sigma}\frac{\sqrt{\psid^{-1}\snr\cdot\left( 1-\latticeLoose\bsnr^{-1}\right)}{ }}{1+\psid^{-1}\snr},
\end{align}
\begin{align}
\label{eq:gammansetting}
  \gamma_n = \sqrt{\frac{1}{\sigma^2_n}\left(\frac{\wt{P}}{\latticeLoose}-\wt{\sigma}^2\right)}, 
\end{align}
where 
\begin{align}
\label{eq:sigmanmod}
\sigma^2_n=
\psid{\snr^{-1}
\left(1+\snr\cdot\psid^{-1}\frac{1-\latticeLoose\cdot\bsnr^{-1}}{1+\psid\latticeLoose\cdot\dsnr^{-1}} \right)^{1-n}}. 
\end{align}
Define: 
\begin{align}
\label{eq:penalty}
\Psi_1 &\dfn 1+\psid\latticeLoose\cdot\dsnr^{-1}  \\ 
\Psi_2 &\dfn \frac{1}{1-\latticeLoose\cdot\bsnr^{-1}} \\
\psid&\dfn 1+\frac{3}{2}Lp'_m\left(N-1-\frac{2}{N}\right)\label{eq:Psi4}\\
\psic &\dfn \exp\left(\tfrac{1}{\snr\cdot \psid^{-1}(\Psi_1 \Psi_2)^{-\frac{N-1}{N}}	
	\Gamma^{-\frac{1}{N}}_{0}\left(\frac{\Pet}{2}\right) -1} \right)
\end{align}

\begin{theorem}\label{thrm:capgap}
For the choice of parameters above, the interactive communication scheme described in Subsection~\ref{ourscheme} achieves in $N$ rounds an error probability $\Pe\leq\Pet$ and a capacity gap $\Gamma_{\db}^*$ satisfying: 
\begin{align}
  \label{eq:capgapourshceme}
&\Gamma_{\db}^*(\Pet,N)\\ &<  \tfrac{1}{N}\Gamma_{0,\db}\left(\tfrac{\Pet}{2}\right) + \tfrac{N-1}{N}\left(\Psi_{1,\db}  + \Psi_{2,\db}\right) +\psiddb + \psicdb, 
\end{align}
provided that $\psic>1$ (see Remark \ref{rem:psic}). The penalty term $\psid$ is typically negligible
(see Remark \ref{rem:concat}).
\end{theorem}

We prove this theorem is Subsection \ref{sec:proof}.

\begin{corollary}[High $\snr$ behavior]
Let $\dsnr$ and $\Pet$ be fixed. The capacity gap attained by our scheme for $\snr$ large enough, can be approximated by
\begin{align}
\Gamma^*_{\db}(\Pet,N)&\\ \approx& \tfrac{1}{N}\Gamma_{0,\db}\left(\tfrac{\Pet}{2}\right) + 
\tfrac{N-1}{N}\left[1+\latticeLoose{\dsnr}^{-1}\right]_\db. \label{eq:capgapapprox}
\end{align}
\end{corollary}
Note that the first term is roughly the capacity gap of the S-K scheme with noiseless feedback, and that the second term depends only on $\dsnr$.

The following remarks are in order. 

\begin{remark}\label{rem:noiseinsertion}
The penalty term $\Psi_1$ can be attributed to \textit{noise insertion} from the feedback channel to the feedforward channel. As shown below in \eqref{eq:coupledTransmission}, our feedforward transmission can be interpreted as analog transmission with an additional noise term emanating from the feedback channel. The aggregate noise variance grows from $\sigma^2$ to $\Psi_1\sigma^2$, causing a decrease in the signal-to-noise ratio, and a corresponding increase in the capacity gap.
\end{remark}

\begin{remark}\label{rem:Psi2}
The penalty term $\Psi_2$ can be attributed to \textit{power loss}, which is a consequence of the noise insertion discussed above. Recall that Terminal A sends $\alpha\gamma_n\wt{\eps}_n$ which is a (scaled) noisy version of the true estimation error $\eps_n$, and where the noise stems from the feedback channel. Thus, part of the transmission power $P$ is consumed by this noise, leaving only a power $P/\Psi_2 < P$ for the description of $\eps_n$. This in turn reduces the  signal-to-noise ratio and correspondingly increases the capacity gap.
\end{remark}

\begin{remark}\label{rem:concat}
The penalty term $\psid$ can be attributed to the deviation in the feedforward transmission power caused by modulo-aliasing errors. It is important to note that $\psiddb=0\db$ for $N<3$. Moreover, $\psiddb$ is typically negligible for $N\geq 3$ and practical settings of $\Pet$.
This can be seen as follows: Recalling \eqref{eq:Ldef} and using the inverse form of the exponential bound for the $Q$ function, $Q^{-1}(x)\leq \sqrt{-2\ln(2x)}$, yields 
\begin{align}
\psid&\leq 1+\frac{3}{2}
\sqrt{-2\ln\left(\frac{\Pet}{N-1}\right)}\cdot
\frac{\Pet}{2(N-1)}
\left(N-1-\frac{2}{N}\right)\\
&\leq
1+\frac{3}{4}\Pet
\sqrt{-2\ln\left(\frac{\Pet}{N-1}\right)}.
\end{align}
Setting e.g. $\Pet=10^{-6}$ and $N=10$ in the above equation yields $\psiddb\leq 2\cdot 10^{-5}\db$, which is clearly negligible for all practical purposes.
\end{remark}

\begin{remark}
	\label{rem:psic}
	$\psic$ is an additional penalty term, that results from the fact that we consider the capacity gap in terms of $\snr$ ratios, whereas the explicit term arising from the capacity formula is related to $\log{(1+\snr)}$ rather than $\log{(\snr)}$. Note that $\psicdb = \mathrm{O}\left(\snr^{-1}\right)$. Due to simplifying assumptions taken in the bounding technique it is necessary to assume that $\psicdb>0$. The case where $\psicdb<0$ corresponds to settings of very low $\snr$/ low $\dsnr$/ very low $\Pet$ which is a less interesting regime. Note that in this case the capacity gap can still be calculated, but the bound is more cumbersome and is left out. 
\end{remark}

\begin{remark}
Note that there is a ``low $\snr$'' regime (related also to the target error probability or to $\dsnr$), where the loss terms $\Psi_{1,\db}+\Psi_{2,\db}$ are larger than say $\Gamma_{0,\db}\left(\Pet\right)$. In that case, setting $N=1$, namely using an uncoded system with no interaction, is the optimal choice of parameters for our scheme. 
\end{remark}

\subsection{Proof of Theorem~\ref{thrm:capgap} }\label{sec:proof}
In Subsection \ref{s-kbasic} we analyzed the error probability of the S-K scheme with noiseless feedback, relying on the fact that all the noises are jointly Gaussian, including the noise $\varepsilon_N$ experienced by the PAM decoder. To that end, we were able to directly use the error probability analysis of simple PAM over an AWGN channel discussed in Subsection \ref{pambasic}. 

In the noisy feedback case however, the non-linearity of the modulo operations at both terminals induces a non-Gaussian distribution of $\varepsilon_N$. An analysis of the decoding error based on the actual distribution of $\varepsilon_N$ seems involved. Yet, an upper bound can be derived via a simple coupling argument described below. 

Recall that Terminal A computes $\wt{\eps}_n$, a noisy  version of the estimation error at Terminal B, via a modulo operation \eqref{eq:noisy_estimate}. For any $n\in\{1,\ldots, N-1\}$ we define $E_n$ as the event where this computation results in a modulo-aliasing error, i.e.,
\begin{align}
\label{eq:moderror}
E_n\dfn\{ \gamma_n\varepsilon_n+\wt{Z}_n\notin[-\tfrac{d}{2},\tfrac{d}{2})\}.
\end{align}
Furthermore, we define $E_N$ to be the PAM decoding error event:
\begin{align}
E_N=\{ \varepsilon_N\notin[-\tfrac{d_{\mathrm{min}}}{2},\tfrac{d_{\mathrm{min}}}{2})\},
\end{align}
where $d_{\mathrm{min}}$ is the minimal distance of the PAM. The error probability of our scheme is clearly  $\Pe\dfn \Pr(E_N)$. However, as mentioned above, the distribution of  $\varepsilon_N$ is not Gaussian due to the nonlinearity introduced by the modulo operations, which in turn renders $\Pe$ difficult to compute. To circumvent this problem, we  instead consider the following trivial upper bound   
\begin{align}\label{eq:err_union}
\Pe \leq \Pr\left(\bigcup_{n=1}^NE_n \right).
\end{align}
We intuitively expect this bound to be rather tight, since a modulo-aliasing error is very likely to imply a PAM decoding error. It turns out that upper bounding the right-hand-side above is not too difficult, as we now show.

To proceed, we define the \textit{coupled system} as a system that is fed by the exact same message, and experiences the (sample-path) exact same noises. The coupled system differs from the original system in two aspects: first, it does not apply any modulo operations at either of the terminals; second, its feedforward power constraint is set to be $P'={P}/{\psid}$ (where we appropriately define $\snr'\dfn\snr/\psid$). We denote all the signals and events in the coupled system using the same notation as in the original system, but with an additional prime $(')$ symbol, unless the signals are always identical between the systems by construction (e.g. $Z_n$). Clearly, the coupled system can violate the power constraint at Terminal B. However, given the message $W$, all the random variables in the coupled system are jointly Gaussian, and in particular, the estimation errors $\varepsilon_n'$ in that system are Gaussian for $n=1,\ldots,N$. Moreover, it is easy to see that given no modulo-aliasing has occurred up to time $n$, the estimation errors are \textit{sample-path identical} between the systems, i.e. $\varepsilon_n' = \varepsilon_n$. This leads to the following lemma:

\begin{lemma}
\label{lemma:coupling}
For any $N>1$:
\begin{align}
\Pr\left(\bigcup_{n=1}^N E_n \right) =  \Pr\left(\bigcup_{n=1}^N E_n' \right).
\end{align}
 \end{lemma}

\begin{IEEEproof}
Define the event 
\begin{align}\label{eq:Jn}
  J_n \dfn \bigcap_{i=1}^n \overline{E_i}
\end{align}
Let us show by induction that $J_n = J_n'$. For $n=1$, we have 
\begin{align}\label{eq:J1}
  J_1 &= \{\gamma_1\varepsilon_1+\wt{Z}_1\in[-\tfrac{d}{2},\tfrac{d}{2})\} \nonumber \\
  & = \{\gamma_1\varepsilon_1'+\wt{Z}_1\in[-\tfrac{d}{2},\tfrac{d}{2})\} \\
  & = J_1' \nonumber
\end{align}
where~\eqref{eq:J1} follows from the sample path identity. Assuming $J_{k-1}=J_{k-1}'$ and using the sample path identity again, we have  
\begin{align}
  J_k &= \{\gamma_k\varepsilon_k+\wt{Z}_k\in[-\tfrac{d}{2},\tfrac{d}{2})\} \cap  J_{k-1} \\
  & = \{\gamma_k\varepsilon_k'+\wt{Z}_k\in[-\tfrac{d}{2},\tfrac{d}{2})) \} \cap J_{k-1}' \\
  & = J_k' 
\end{align}
By the exact same argument (replacing $\in$ with $\notin$) we clearly have that $J_{n-1}\cap E_n = J_{n-1}'\cap E_n'$. Thus we can write 
\begin{align}
\Pr\left(\bigcup_{n=1}^N E_n \right) &=  \Pr(E_1) +  \sum_{n=2}^{N}\Pr\left(\bigcap_{i=1}^{n-1}\overline{E_i}\cap E_n\right) \\ 
&= \Pr(\overline{J_1}) + \sum_{n=2}^{N}\Pr\left(J_{n-1}\cap E_n\right) \\ 
&= \Pr(\overline{J'_1}) + \sum_{n=2}^{N}\Pr\left(J_{n-1}'\cap E_n'\right) \\ 
&= \Pr\left(\bigcup_{n=1}^N E_n' \right)
\end{align}
\end{IEEEproof}
Combining the above with \eqref{eq:err_union} and applying the union bound in the coupled system, we obtain
\begin{align}\label{eq:union_bound}
\Pe\leq\sum_{n=1}^{N}\Pr\left(E_n'\right) .
\end{align}
Thus, we can now upper bound the error probability by calculating probabilities in the coupled system, which involves only scalar Gaussian densities and significantly simplifies the analysis. 

Let us begin by calculating the scheme parameters, and then use them to calculate the capacity gap. As mentioned in Subsection~\ref{sec:main-res}, the target error probability is set to $\Pet$. Let us set $\Pr(E'_N)=\frac{\Pet}{2}$, and $\Pr(E'_1) =\cdots = \Pr(E'_{n-1}) \dfn p'_m$, where $p'_m$ is given in \eqref{eq:pem}, and calculate the corresponding $L$. Recalling the definition of the event $E'_n$ in \eqref{eq:moderror} and that $d=\sqrt{12\wt{P}}$, and since $\gamma_n\varepsilon'_n+\wt{Z}_n$ is Gaussian, we have that 
\begin{align}
\label{eq:pm}
p'_m=2Q\left(\sqrt{\frac{3\wt{P}}{\Expt (\gamma_n\varepsilon'_n+\wt{Z}_n)^2}}\right).
\end{align}
Using the definition of $\latticeLoose$ in~\eqref{eq:latticecLooseDef}, where here $\sigma^2_\eps=\sigma^2_n=\Expt(\eps'_n)^2$ yields
\begin{align}\label{eq:sigmaAndL}
  \Expt (\gamma_n\varepsilon'_n+\wt{Z}_n)^2 = \gamma_n^2\sigma^2_n+\wt{\sigma}^2 = \frac{\wt{P}}{\latticeLoose}.
\end{align}
First note that solving this equation for $\gamma_n$ yields \eqref{eq:gammansetting}.
Also note that combining the above two equations gives us the following setting of $L$:
\begin{align}
\label{eq:frack}
\latticeLoose =  \tfrac{1}{3}\left[Q^{-1}\left(\frac{p'_m}{2}\right)\right]^{2}.
\end{align}
%

With $L$ in hand, we can easily choose $\alpha$ so that the power constraint at Terminal A is satisfied with equality. Namely  
\begin{align}
P'&=\Expt (X'_{n+1})^2
=\Expt(\alpha\gamma_n\wt{\varepsilon}_n)^2
\\&=\alpha^2\Expt(\gamma_n\wt{\varepsilon}_n)^2
=\alpha^2\Expt (\gamma_n\varepsilon'_n+\wt{Z}_n)^2
. 
\end{align}
From \eqref{eq:sigmaAndL} it follows that:
\begin{align}
\label{eq:alphL}
\alpha=\sqrt{\latticeLoose\frac{P'}{\wt{P}}}.
\end{align}

The parameter $\beta_n$ determines the evolution of the estimation error. 
The linear estimate of $\varepsilon'_n$: $\wh{\varepsilon}'_n=\beta_{n+1}Y'_{n+1}$, is the optimal estimate in the coupled system, in which $\varepsilon'_n$ and $Y'_{n+1}$ are jointly Gaussian. We would thus like to minimize $\Expt\left(\eps'_n-\wh{\eps}'_n\right)^2$. Recalling the input-output relation of the feedforward channel and using \eqref{eq:alphL} we obtain
\begin{align}
\label{eq:Yn1}
Y'_{n+1}&=\sqrt{\tfrac{\latticeLoose P'}{\wt{P}}}\left(\gamma_n\varepsilon'_n+\wt{Z}_n\right)+Z_{n+1} 
\end{align}
and solving the optimization for $\beta_n$ yields:
\begin{align} 
\beta_{n+1}=\frac{\sigma_n}{\sigma}\frac{\sqrt{\snr'\cdot\left( 1-\latticeLoose\bsnr^{-1}\right)}{ }}{1+\snr'},
\end{align}
Noting that $\eps'_{n+1} = \eps'_n-\wh{\eps}'_n$ and computing the MMSE for the optimal choice of $\beta_{n+1}$ above, we obtain a recursive formula for $\sigma_n^2$ and $\snr_n$:

\begin{align}
\snr_n &= \frac{1}{\sigma^2_n}\\
&=\snr'\cdot\left(1+\snr'\cdot\frac{1-\latticeLoose\cdot\bsnr^{-1}}{1+\psid\latticeLoose\cdot\dsnr^{-1}} \right)^{n-1} \\ 
&=
\snr'\cdot\left(1+\snr'\cdot\Psi_1\Psi_2 \right)^{n-1}.\label{eq:snreqmod}
\end{align}

It is possible to give a different and modular interpretation for \eqref{eq:snreqmod}. Let us rewrite \eqref{eq:Yn1}:
\begin{align}
\label{eq:coupledTransmission}
Y'_{n+1}=\sqrt{\tfrac{\latticeLoose P'}{\wt{P}}}\gamma_n\eps'_n+\sqrt{\tfrac{\latticeLoose P'}{\wt{P}}}\wt{Z}_n+Z_{n+1} 
\end{align}
This equation can be regarded as a JSCC problem designated for the transmission of $\eps'_n$ over an AWGN channel. The effective noise of this AWGN channel is $\sqrt{\tfrac{\latticeLoose P'}{\wt{P}}}\wt{Z}_n+Z_{n+1}$. Some algebra shows that the variance of this noise is $\Psi_1\sigma^2$ where $\Psi_1$ is defined in \eqref{eq:penalty}. We call this phenomenon \textit{noise insertion}, as previously mentioned in Remark~\ref{rem:noiseinsertion}. A consequence of this phenomenon is that part of the transmission power is now consumed by the noise element related to $\wt{Z}_n$. Subtracting this penalty from $P'$ shows that the part of transmission power used for the description of $\eps'_n$ is  $P'/\Psi_2$. We call this phenomena \textit{power loss} as previously mentioned in Remark~\ref{rem:Psi2}. So, all in all, the $\snr'$ of the channel describe by \eqref{eq:coupledTransmission} is $\Psi_1\Psi_2\cdot \snr'$. Using this fact together with the $\snr$ evolution of the S-K scheme \eqref{eq:snreq}, and noting that the noise insertion and power loss effects only occur after the second round, we obtain \eqref{eq:snreqmod}.

We are now left with calculating the capacity gap. 
Let us find the condition that guarantees our scheme operates within the target error probability requirement. Rewriting \eqref{eq:union_bound} we get 
\begin{align}
\Pe&\leq\sum_{n=1}^{N}
\Pr\left( E_n'\right)\\
&\leq(N-1)p'_m+\Pr\left(E'_N\right)\\
&\leq\frac{\Pet}{2}+\Pr\left(E'_N\right)\\
&\leq\frac{\Pet}{2}+ 2Q\left(\sqrt{\frac{3\snr_N}{2^{2NR}-1}} \right) \label{eq:delta2Q}, 
\end{align}
hence by setting  
\begin{align}\label{eq:Pt}
Q\left(\sqrt{\frac{3\snr_N}{2^{2NR}-1}}\right)=\frac{\Pet}{4}
\end{align}
we obtain that $\Pe\leq \Pet$ as desired. We are now in a position to derive a lower bound on the capacity gap attained by our scheme: we can rearrange~\eqref{eq:Pt} to obtain a lower bound on $R$, use the expression~\eqref{eq:snreqmod} for $\snr_N$, and plug this into the definition of the capacity gap~\eqref{eq:capGapDef}. This yields~\eqref{eq:capgapourshceme}, where $\psicdb$ is a remainder term obtained by pedestrian manipulations using the inequality  $-\ln(1-x)\leq \frac{x}{1-x}$ for $x\in(0,1)$. Note that the result was obtained for the specific choice $p'_m = \frac{\Pet}{2(N-1)}$ of the modulo-aliasing error. In general, reducing $p_m$ increases $\latticeLoose$ which in turn decreases $\snr_N$, and hence increases the second addend on the right-hand-side of \eqref{eq:delta2Q}, resulting in a trade-off that could potentially be further optimized. 

We are now left with specifying the relation between $P'$ and $P$, captured by $\psid$. We start by defining the following sequence of events:
\begin{align}
A_n \dfn \{X_n=X'_n\}\quad\text{for } n=1,...,N. 
\end{align}
Using this definition, we can write 
\begin{align}
&\Expt(X_n^2) = \Pr(A_n)\Expt\left(X_n^2\mid A_n\right)
+\Pr(\overline{A_n})\Expt\left(X_n^2\mid \overline{A_n}\right)\\
&\Expt(X'_n)^2 = \Pr(A_n)\Expt\left((X'_n)^2\mid A_n\right)
+\Pr(\overline{A_n})\Expt\left((X'_n)^2\mid \overline{A_n}\right).
\end{align}
Combining these expressions, while noting that $\Expt(X'_n)^2=P'$ and that $\Expt\left((X'_n)^2\mid A_n\right)=\Expt\left(X_n^2\mid A_n\right)$, we obtain:
\begin{align}
\Expt(X_n^2) &= P'+\Pr(\overline{A_n})\cdot\Expt\left(X_n^2-(X'_n)^2\mid \overline{A_n}\right) \\ 
& \leq P'+\Pr(\overline{A_n})\cdot\Expt\left(X_n^2\mid \overline{A_n}\right) \label{eq:XnPower}
\end{align}
Let us now calculate \eqref{eq:XnPower} separately for $n<3$ and $n\geq 3$. For $n=1$ we have $X_1=X'_1$ by construction, hence  $\Pr(\overline{A_1})=0$ and thus $\Expt(X_1)^2=P'$. For $n=2$ we have by \eqref{eq:epsntilde}:
\begin{align}
X_n&=\alpha \moduloOp{ \gamma_1\eps_1+\wt{Z}_1} \\
X'_n&=\alpha \left(\gamma_1\eps_1+\wt{Z}_1\right).
\end{align}
By definition of the scalar modulo operation it holds that $|\moduloOp{x}|\leq |x|$, and therefore $|X_n|\leq |X'_n|$. , which implies that $\Expt(X_n)^2\leq P'$. We are now left with the more general case of $n\geq 3$. We have that  
\begin{align}
\Expt\left(X_n^2 \mid \overline{A_n}\right)&\leq\Expt\left(\left(\alpha \moduloOp{ \gamma_n\eps_n+\wt{Z}_n}\right)^2\mid \overline{A_n}\right) \\
&= \left(\alpha \frac{d}{2}\right)^2\\
&= \left(\sqrt{L\frac{P'}{\wt{P}}} \frac{\sqrt{12\wt{P}}}{2}\right)^2\\
&=3LP'
\end{align}
Plugging the above in \eqref{eq:XnPower} yields
\begin{align}
\Expt(X_n)^2 \leq P'\left(1+3\Pr(\overline{A_n})L\right)
\end{align}
We are left with bounding $\Pr(\overline{A_n})$ for $n\geq 3$. To that end, note that $X_n\neq X'_n$ implies that at least one  modulo error occurred up to time instant $n-1$, with probability one. Therefore:
\begin{align}
\Pr\left(\overline{A_n}\right)\leq
\Pr\left(\bigcup_{i=1}^{n-1}{E}_i\right)\leq \sum_{i=1}^{n-1}\Pr\left({E}_i\right)=(n-1)p'_m
\end{align}

Collecting the upper bounds on $\Expt(X_n^2)$ for all $n=1,..,N$, we obtain the following upper bound on the average power consumed by our scheme: 
\begin{align}
P&=\frac{1}{N}\sum_{n=1}^N\mathbb{E}(X_n)^2\\ 
&\leq \frac{P'}{N}\left(2+\sum_{n=3}^{N}\left(1+3L(n-1)p'_m\right)\right)\\
&=\frac{P'}{N}\left(N+\frac{3(N-2)(N+1)}{2}Lp'_m\right)\\ 
&= P'\left(1+\frac{3}{2}Lp'_m\left(N-1-\frac{2}{N}\right)\right).
\end{align}
Hence, setting $P'= P / \psid$ where $\psid$ is given by \eqref{eq:Psi4} obeys the feedforward transmission power constraint. This concludes the proof.
\subsection{Numerical Results}
\label{results}
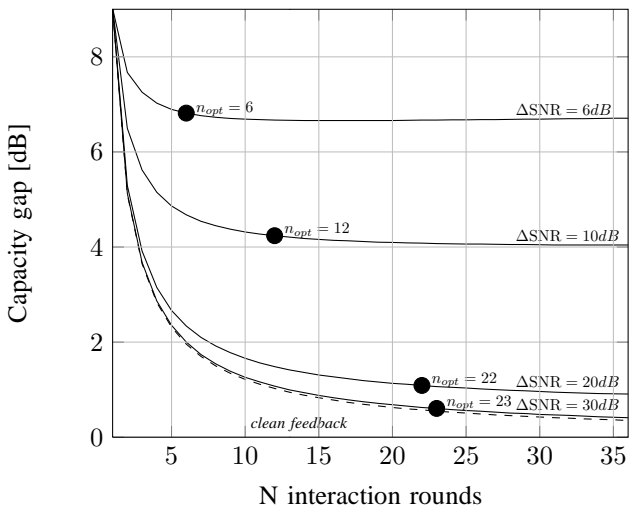
\begin{figure}
\centering
%
%
%
%
\begin{tikzpicture}

\begin{axis}[
xlabel={N interaction rounds},
ylabel={Capacity gap [dB]},
xmin=1, xmax=36,
ymin=0, ymax=9,
axis on top,
xmajorgrids,
ymajorgrids
]
\addplot [black, dashed]
coordinates {
(1,9.01787449938529)
(2,5.107421875)
(3,3.6474609375)
(4,2.83203125)
(5,2.314453125)
(6,1.9580078125)
(7,1.69921875)
(8,1.4990234375)
(9,1.3427734375)
(10,1.2158203125)
(11,1.11328125)
(12,1.025390625)
(13,0.947265625)
(14,0.8837890625)
(15,0.830078125)
(16,0.78125)
(17,0.732421875)
(18,0.693359375)
(19,0.6591796875)
(20,0.625)
(21,0.595703125)
(22,0.576171875)
(23,0.546875)
(24,0.52734375)
(25,0.5078125)
(26,0.48828125)
(27,0.46875)
(28,0.4541015625)
(29,0.439453125)
(30,0.4248046875)
(31,0.41015625)
(32,0.400390625)
(33,0.3857421875)
(34,0.3759765625)
(35,0.3662109375)
(36,0.3515625)

};
\addplot [black]
coordinates {
(1,9.01787449938529)
(2,7.6708984375)
(3,7.255859375)
(4,7.0263671875)
(5,6.89453125)
(6,6.81640625)
(7,6.7626953125)
(8,6.728515625)
(9,6.7041015625)
(10,6.689453125)
(11,6.6796875)
(12,6.669921875)
(13,6.6650390625)
(14,6.66015625)
(15,6.66015625)
(16,6.66015625)
(17,6.66015625)
(18,6.66015625)
(19,6.66015625)
(20,6.66015625)
(21,6.6650390625)
(22,6.669921875)
(23,6.669921875)
(24,6.6748046875)
(25,6.6748046875)
(26,6.6796875)
(27,6.6796875)
(28,6.6845703125)
(29,6.689453125)
(30,6.689453125)
(31,6.6943359375)
(32,6.69921875)
(33,6.69921875)
(34,6.7041015625)
(35,6.708984375)
(36,6.708984375)

};
\addplot [black, mark=*, mark size=3, only marks]
coordinates {
(6,6.81640625)

};
\addplot [black]
coordinates {
(1,9.01787449938529)
(2,6.494140625)
(3,5.6201171875)
(4,5.1513671875)
(5,4.86328125)
(6,4.677734375)
(7,4.541015625)
(8,4.4482421875)
(9,4.375)
(10,4.31640625)
(11,4.2724609375)
(12,4.23828125)
(13,4.208984375)
(14,4.1796875)
(15,4.16015625)
(16,4.140625)
(17,4.130859375)
(18,4.1162109375)
(19,4.1015625)
(20,4.0966796875)
(21,4.0869140625)
(22,4.08203125)
(23,4.072265625)
(24,4.0673828125)
(25,4.0625)
(26,4.0625)
(27,4.0576171875)
(28,4.052734375)
(29,4.052734375)
(30,4.0478515625)
(31,4.04296875)
(32,4.04296875)
(33,4.04296875)
(34,4.04296875)
(35,4.04296875)
(36,4.04296875)

};
\addplot [black, mark=*, mark size=3, only marks]
coordinates {
(12,4.23828125)

};
\addplot [black]
coordinates {
(1,9.01787449938529)
(2,5.29296875)
(3,3.916015625)
(4,3.1494140625)
(5,2.666015625)
(6,2.3388671875)
(7,2.099609375)
(8,1.9189453125)
(9,1.77734375)
(10,1.66015625)
(11,1.5625)
(12,1.484375)
(13,1.416015625)
(14,1.3623046875)
(15,1.30859375)
(16,1.26953125)
(17,1.23046875)
(18,1.19140625)
(19,1.162109375)
(20,1.1328125)
(21,1.11328125)
(22,1.0888671875)
(23,1.064453125)
(24,1.0498046875)
(25,1.03515625)
(26,1.015625)
(27,1.0009765625)
(28,0.986328125)
(29,0.9765625)
(30,0.9619140625)
(31,0.947265625)
(32,0.9375)
(33,0.927734375)
(34,0.91796875)
(35,0.9130859375)
(36,0.908203125)

};
\addplot [black, mark=*, mark size=3, only marks]
coordinates {
(22,1.0888671875)

};
\addplot [black]
coordinates {
(1,9.01787449938529)
(2,5.126953125)
(3,3.6767578125)
(4,2.8662109375)
(5,2.353515625)
(6,2.001953125)
(7,1.73828125)
(8,1.54296875)
(9,1.38671875)
(10,1.259765625)
(11,1.162109375)
(12,1.07421875)
(13,0.99609375)
(14,0.9375)
(15,0.87890625)
(16,0.830078125)
(17,0.7861328125)
(18,0.7470703125)
(19,0.712890625)
(20,0.68359375)
(21,0.654296875)
(22,0.625)
(23,0.60546875)
(24,0.5810546875)
(25,0.5615234375)
(26,0.546875)
(27,0.52734375)
(28,0.5078125)
(29,0.498046875)
(30,0.478515625)
(31,0.46875)
(32,0.458984375)
(33,0.4443359375)
(34,0.4296875)
(35,0.419921875)
(36,0.41015625)

};
\addplot [black, mark=*, mark size=3, only marks]
coordinates {
(23,0.60546875)

};
\path [draw=black, fill opacity=0] (axis cs:13,9)--(axis cs:13,9);

\path [draw=black, fill opacity=0] (axis cs:36,13)--(axis cs:36,13);

\path [draw=black, fill opacity=0] (axis cs:13,0)--(axis cs:13,0);

\path [draw=black, fill opacity=0] (axis cs:1,13)--(axis cs:1,13);

\node at (axis cs:10,0.2)[
  scale=0.6,
  anchor=base west,
  text=black,
  rotate=0.0
]{\itshape clean feedback};
\node at (axis cs:28,6.769453125)[
  scale=0.6,
  anchor=base west,
  text=black,
  rotate=0.0
]{ $\Delta\mathrm{SNR}=6dB$};
\node at (axis cs:6.3,6.8426953125)[
  scale=0.6,
  anchor=base west,
  text=black,
  rotate=0.0
]{ $n_{opt}=6$};
\node at (axis cs:28,4.132734375)[
  scale=0.6,
  anchor=base west,
  text=black,
  rotate=0.0
]{ $\Delta\mathrm{SNR}=10dB$};
\node at (axis cs:12.3,4.288984375)[
  scale=0.6,
  anchor=base west,
  text=black,
  rotate=0.0
]{ $n_{opt}=12$};
\node at (axis cs:28,1.0565625)[
  scale=0.6,
  anchor=base west,
  text=black,
  rotate=0.0
]{ $\Delta\mathrm{SNR}=20dB$};
\node at (axis cs:22.3,1.144453125)[
  scale=0.6,
  anchor=base west,
  text=black,
  rotate=0.0
]{ $n_{opt}=22$};
\node at (axis cs:28,0.578046875)[
  scale=0.6,
  anchor=base west,
  text=black,
  rotate=0.0
]{ $\Delta\mathrm{SNR}=30dB$};
\node at (axis cs:23.3,0.6610546875)[
  scale=0.6,
  anchor=base west,
  text=black,
  rotate=0.0
]{ $n_{opt}=23$};
\end{axis}

\end{tikzpicture}
\caption{\label{fig:resultsFigR1}
The capacity gap as function of the iterations and $\dsnr$ for a target rate $R=1$ (low $\snr$), 
and target error probability $p_t=10^{-6}$}
\end{figure}

\begin{figure}
\centering
%
%
%
%
\begin{tikzpicture}

\begin{axis}[
xlabel={N interaction rounds},
ylabel={Capacity gap [dB]},
xmin=1, xmax=36,
ymin=0, ymax=9,
axis on top,
xmajorgrids,
ymajorgrids
]
\addplot [black, dashed]
coordinates {
(1,9.01787449938529)
(2,4.228515625)
(3,2.822265625)
(4,2.119140625)
(5,1.6943359375)
(6,1.416015625)
(7,1.2109375)
(8,1.0595703125)
(9,0.9423828125)
(10,0.849609375)
(11,0.771484375)
(12,0.7080078125)
(13,0.654296875)
(14,0.60546875)
(15,0.56640625)
(16,0.5322265625)
(17,0.5029296875)
(18,0.4736328125)
(19,0.44921875)
(20,0.4248046875)
(21,0.4052734375)
(22,0.390625)
(23,0.37109375)
(24,0.3564453125)
(25,0.341796875)
(26,0.3271484375)
(27,0.3173828125)
(28,0.302734375)
(29,0.29296875)
(30,0.283203125)
(31,0.2734375)
(32,0.2685546875)
(33,0.2587890625)
(34,0.25390625)
(35,0.244140625)
(36,0.2392578125)

};
\addplot [black]
coordinates {
(1,9.01787449938529)
(2,7.8173828125)
(3,7.734375)
(4,7.7197265625)
(5,7.7294921875)
(6,7.744140625)
(7,7.763671875)
(8,7.783203125)
(9,7.802734375)
(10,7.8173828125)
(11,7.83203125)
(12,7.8515625)
(13,7.861328125)
(14,7.8759765625)
(15,7.890625)
(16,7.9052734375)
(17,7.9150390625)
(18,7.9296875)
(19,7.939453125)
(20,7.94921875)
(21,7.958984375)
(22,7.96875)
(23,7.978515625)
(24,7.9833984375)
(25,7.9931640625)
(26,8.0029296875)
(27,8.0078125)
(28,8.017578125)
(29,8.02734375)
(30,8.0322265625)
(31,8.037109375)
(32,8.046875)
(33,8.0517578125)
(34,8.056640625)
(35,8.06640625)
(36,8.0712890625)

};
\addplot [black, mark=*, mark size=3, only marks]
coordinates {
(4,7.7197265625)

};
\addplot [black]
coordinates {
(1,9.01787449938529)
(2,6.69921875)
(3,6.2158203125)
(4,6.005859375)
(5,5.888671875)
(6,5.8203125)
(7,5.78125)
(8,5.751953125)
(9,5.732421875)
(10,5.72265625)
(11,5.712890625)
(12,5.7080078125)
(13,5.703125)
(14,5.703125)
(15,5.703125)
(16,5.703125)
(17,5.7080078125)
(18,5.7080078125)
(19,5.712890625)
(20,5.712890625)
(21,5.712890625)
(22,5.7177734375)
(23,5.72265625)
(24,5.72265625)
(25,5.7275390625)
(26,5.732421875)
(27,5.732421875)
(28,5.7373046875)
(29,5.7421875)
(30,5.7421875)
(31,5.7470703125)
(32,5.751953125)
(33,5.751953125)
(34,5.7568359375)
(35,5.76171875)
(36,5.76171875)

};
\addplot [black, mark=*, mark size=3, only marks]
coordinates {
(5,5.888671875)

};
\addplot [black]
coordinates {
(1,9.01787449938529)
(2,5.556640625)
(3,4.6630859375)
(4,4.2333984375)
(5,3.984375)
(6,3.828125)
(7,3.7158203125)
(8,3.6376953125)
(9,3.5791015625)
(10,3.53515625)
(11,3.49609375)
(12,3.466796875)
(13,3.4423828125)
(14,3.427734375)
(15,3.408203125)
(16,3.3935546875)
(17,3.3837890625)
(18,3.3740234375)
(19,3.3642578125)
(20,3.359375)
(21,3.3544921875)
(22,3.349609375)
(23,3.3447265625)
(24,3.33984375)
(25,3.3349609375)
(26,3.3349609375)
(27,3.330078125)
(28,3.330078125)
(29,3.330078125)
(30,3.3251953125)
(31,3.3251953125)
(32,3.3251953125)
(33,3.3203125)
(34,3.3203125)
(35,3.3203125)
(36,3.3203125)

};
\addplot [black, mark=*, mark size=3, only marks]
coordinates {
(11,3.49609375)

};
\addplot [black]
coordinates {
(1,9.01787449938529)
(2,4.404296875)
(3,3.06640625)
(4,2.40234375)
(5,2.0068359375)
(6,1.73828125)
(7,1.552734375)
(8,1.4111328125)
(9,1.3037109375)
(10,1.2158203125)
(11,1.1474609375)
(12,1.0888671875)
(13,1.0400390625)
(14,0.99609375)
(15,0.95703125)
(16,0.927734375)
(17,0.8984375)
(18,0.8740234375)
(19,0.8544921875)
(20,0.830078125)
(21,0.8154296875)
(22,0.80078125)
(23,0.78125)
(24,0.771484375)
(25,0.7568359375)
(26,0.7470703125)
(27,0.7373046875)
(28,0.7275390625)
(29,0.7177734375)
(30,0.7080078125)
(31,0.703125)
(32,0.693359375)
(33,0.6884765625)
(34,0.68359375)
(35,0.673828125)
(36,0.6689453125)

};
\addplot [black, mark=*, mark size=3, only marks]
coordinates {
(19,0.8544921875)

};
\addplot [black]
coordinates {
(1,9.01787449938529)
(2,4.248046875)
(3,2.8466796875)
(4,2.1484375)
(5,1.728515625)
(6,1.4453125)
(7,1.25)
(8,1.0986328125)
(9,0.9814453125)
(10,0.888671875)
(11,0.810546875)
(12,0.7470703125)
(13,0.693359375)
(14,0.6494140625)
(15,0.60546875)
(16,0.576171875)
(17,0.5419921875)
(18,0.517578125)
(19,0.48828125)
(20,0.46875)
(21,0.44921875)
(22,0.4296875)
(23,0.4150390625)
(24,0.400390625)
(25,0.3857421875)
(26,0.37109375)
(27,0.361328125)
(28,0.3515625)
(29,0.341796875)
(30,0.33203125)
(31,0.322265625)
(32,0.3125)
(33,0.302734375)
(34,0.2978515625)
(35,0.29296875)
(36,0.283203125)

};
\addplot [black, mark=*, mark size=3, only marks]
coordinates {
(20,0.46875)

};
\path [draw=black, fill opacity=0] (axis cs:13,9)--(axis cs:13,9);

\path [draw=black, fill opacity=0] (axis cs:36,13)--(axis cs:36,13);

\path [draw=black, fill opacity=0] (axis cs:13,0)--(axis cs:13,0);

\path [draw=black, fill opacity=0] (axis cs:1,13)--(axis cs:1,13);

\node at (axis cs:10,0.2)[
  scale=0.6,
  anchor=base west,
  text=black,
  rotate=0.0
]{\itshape clean feedback};
\node at (axis cs:28,8.10734375)[
  scale=0.6,
  anchor=base west,
  text=black,
  rotate=0.0
]{ $\Delta\mathrm{SNR}=3dB$};
\node at (axis cs:4.3,7.8094921875)[
  scale=0.6,
  anchor=base west,
  text=black,
  rotate=0.0
]{ $n_{opt}=4$};
\node at (axis cs:28,5.8221875)[
  scale=0.6,
  anchor=base west,
  text=black,
  rotate=0.0
]{ $\Delta\mathrm{SNR}=6dB$};
\node at (axis cs:5.3,5.9003125)[
  scale=0.6,
  anchor=base west,
  text=black,
  rotate=0.0
]{ $n_{opt}=5$};
\node at (axis cs:28,3.410078125)[
  scale=0.6,
  anchor=base west,
  text=black,
  rotate=0.0
]{ $\Delta\mathrm{SNR}=10dB$};
\node at (axis cs:11.3,3.546796875)[
  scale=0.6,
  anchor=base west,
  text=black,
  rotate=0.0
]{ $n_{opt}=11$};
\node at (axis cs:28,0.7977734375)[
  scale=0.6,
  anchor=base west,
  text=black,
  rotate=0.0
]{ $\Delta\mathrm{SNR}=20dB$};
\node at (axis cs:19.3,0.910078125)[
  scale=0.6,
  anchor=base west,
  text=black,
  rotate=0.0
]{ $n_{opt}=19$};
\node at (axis cs:28,0.421796875)[
  scale=0.6,
  anchor=base west,
  text=black,
  rotate=0.0
]{ $\Delta\mathrm{SNR}=30dB$};
\node at (axis cs:20.3,0.52921875)[
  scale=0.6,
  anchor=base west,
  text=black,
  rotate=0.0
]{ $n_{opt}=20$};
\end{axis}

\end{tikzpicture}
\caption{\label{fig:resultsFigR4}
The capacity gap as function of the iterations and $\dsnr$ for a target rate $R\geq4$ (high $\snr$),
and target error probability $p_t=10^{-6}$}
\end{figure}

The behavior of the capacity gap for our scheme as a function of the number of interaction rounds and $\dsnr$ is depicted in Fig.~\ref{fig:resultsFigR1} and 
Fig.~\ref{fig:resultsFigR4}, for high $\snr$ and low $\snr$ setups. In both figures we plotted the capacity gap, for a target rate $R$ and a target error probability $p_e=10^{-6}$, where the $\snr$ is found by numerically solving 
$R=\frac{1}{2N}\log\left(1+\frac{\snr_N}{\Gamma_0(\Pet/2)}\right)$ and the capacity gap is calculated by definition. It can be seen that the higher the $\dsnr$, the smaller the capacity gap, where at $\dsnr=30\db$ we virtually obtain the noiseless feedback performance. The points marked $n_{opt}$ are those for which the capacity gap is less than $0.2\db$ above the minimal value attained. In Fig.~\ref{fig:resultsFigR1} the rate was set to $R=1$, and it can be seen that $\dsnr=10\db$ reduces the capacity gap to $4.2\db$ in 12 iterations, and $\dsnr=20\db$ reduces the capacity gap to $1.1\db$ in 22 iterations. In Fig.~\ref{fig:resultsFigR4}, the rate was set to $R=4$, and it can be seen that $\dsnr=10\db$ reduces the capacity gap to $3.5\db$ in 11 iterations, and $\dsnr=20\db$ reduces the capacity gap to $0.8\db$ in 19 iterations. Observing \eqref{eq:capgapapprox} we can see that for high $\snr$ the result is only a function of $\dsnr$, thus does not depend on the target rate or the base $\snr$. 

\subsection{Notes on Implementation}\label{sec:implem}
The scheme described in this section is both simple and practical, as opposed to its noiseless feedback counterparts that break down in the presence of feedback noise. This provides impetus for further discussing implementation related aspects. The following conditions should be met for our results to carry merit: 
\begin{enumerate}[1)]
\item \textit{Information asymmetry}: Terminal A has substantially more information to convey than Terminal B.

\item \textit{$\snr$ asymmetry}: The $\snr$ of the feedforward channel is lower than the $\snr$ of the feedback channel. This can happen due to differences in power constraints (e.g. when Terminal A is battery operated and Terminal B is connected to the power grid), path losses, or noise/interference asymmetry. 

\item \textit{Complexity/delay constraints:} There are severe complexity or delay constraint at Terminal A.

\item \textit{Two-way signaling:} Our scheme assumes sample-wise feedback. The communication system should therefore be full duplex where both terminals have virtually the same signaling rate; hence, the terminals split the bandwidth between them even though only Terminal A is transmitting information. This situation can sometimes be inherent to the system, but should otherwise be tested against the (non-interactive) solution where the entire bandwidth is allocated to Terminal A. This choice of forward vs. feedback bandwidth allocation yields a system trade-off that is $\snr$ dependent: Terminal A can use our scheme and achieve a rate of $C(\snr_{dB}-\Gamma^*_{dB})$ (where $C(\cdot)$ is the Shannon capacity function), or alternatively employ non-interactive codes over the full forward--feedback bandwidth. The latter option doubles the forward signaling rate but also incurs a $3\db$ loss in $\snr$ and a potentially larger capacity gap $\Gamma_\db^\dagger$, resulting in a rate of $2C(\snr_{dB}-3\db -\Gamma_\db^\dagger)$. It can therefore be seen that our solution is generally better for low enough $\snr$. For instance, for $\Pet=10^{-6}$ and $\dsnr > 30\db$ our scheme outperforms (with comparable complexity and delay) full bandwidth uncoded PAM for any $\snr< 23dB$, and outperforms (with significantly smaller complexity and delay) full bandwidth non-feedback codes with $\Gamma_\db^\dagger = 3\db$ for any $\snr < 9\db$.  
\item \textit{Bandwidth splitting}: We have tacitly assumed that the bandwidth is equally split between the feedforward and feedback channels. We note that a non-equal splitting of the bandwidth can also be handled. Standard techniques for trading bandwidth with $\snr$ can be used in order to facilitate the use of our scheme in these scenarios. 
\end{enumerate}

The use of very large PAM constellations, whose size is exponential in the product of rate and interaction rounds, seemingly requires extremely low noise and distortion at the digital and analog circuits in Terminal A. This may appear to impose a major implementation obstacle. Fortunately, this is not the case. The full resolution implied by the constellation size is by construction confined \textit{only} to the original message $\Theta$ and the final estimate $\wh{\Theta}_N$; the transmitted and received signals in the course of interaction can be safely quantized at a resolution determined only by the channel noise (and \textit{not} the final estimation noise), as in commonplace communication systems. Figuratively speaking, the source bits are gradually \textit{revealed} along the interaction process, where the number of bits revealed in every round is determined by the channel $\snr$. 

Another important implementation issue is sensitivity to model assumptions. We have successfully verified in simulation the robustness of the proposed scheme in several reasonable scenarios including correlative noise, excess quantization noise, and multiplicative channel estimation noise. The universality of the scheme and its performance for a wider range of models remains to be further investigated. 

\subsection{Discussion}\label{sec:discussion}
Note that so far we have limited our discussion to the PAM symbol. The bit-error rate is in fact lower, since an error in PAM decoding affects only a single bit with high probability \eqref{eq:pb}, assuming Gray labeling. However, note that the modulo-aliasing error will typically result in many erroneous bits, and hence optimizing the bit error rate does not yield a major improvement over its upper bound. Further fine-tuning of the scheme can be obtained by non-uniform power allocation over interaction rounds in both Terminal A and B; in particular, we note that Terminal B is silent in the last round, which can be trivially exploited. 

We note again that for any choice of $\snr$ and $\dsnr$, the error probability attained by our scheme cannot be made to vanish with the number interaction rounds while maintaining a non-zero rate, as in the noiseless feedback S-K scheme case. The reason is that~\eqref{eq:latticecLooseDef} implies that $\latticeLoose < \bsnr$, which in turn by~\eqref{eq:pmodscalar} imposes a lower bound on the attainable error probability, dictated by the probability of modulo-aliasing of the feedback noise. Equivalently, one cannot get arbitrarily close to capacity for a given target error probability, since increasing the number of iterations improves $\snr_N$ and reduces the PAM decoding error term, but at the same time increases the modulo-aliasing error term in \eqref{eq:delta2Q}. Hence, our scheme is not \textit{capacity achieving} in the usual sense. However, it can get close to capacity in the sense of reducing the capacity gap using a very short block length, typically $N\approx 20$ in the examples presented.  To the best of our knowledge, state-of-the-art (non-interactive) block codes require a block length typically larger by two orders of magnitude to reach the same gap at the same error probability. Consequently, the encoding delay of our scheme is markedly lower than that of these competing schemes. Alternatively, compared to a minimal delay uncoded system under the same error probability, our scheme operates at a much lower capacity gap for a wide regime of settings, and hence can be significantly more power efficient. 

Another important issue is that of encoding and decoding complexity. Our proposed scheme applies only two multiplications and one modulo operation at each terminal in each interaction round. This is significantly lower than the encoding/decoding complexity of good block codes, even if non-optimal methods such as iterative decoding are employed. 

\section{Improving Reliability\label{sec:reliability}}
In this section we describe two asymptotic versions of our simple scheme that are aimed at improving reliability at the cost of increased complexity and delay. These scheme are shown to outperform their non-feedback counterparts in the error exponent sense, when $\dsnr$ is large enough. The first scheme is based on replacing the scalar PAM signaling with a random block code, and replacing the scalar modulo operation with a modulo-lattice operation. The error analysis of both the modulo error and the decoder decision error is performed in the coupled system as before, but are now concerned with error exponents instead of the scalar $Q$-function. The second scheme is based on concatenated coding with the simple interaction scheme as an inner code and a random block outer code.
Classical error exponent results are given in Subsection~\ref{sec:block-codes-error}. JSCC with lattices is discussed in Subsection~\ref{sec:jscc-with-side}. The first scheme is introduced in Subsection~\ref{sec:description-scheme}. A lower bound on its error exponent is given in Subsection~\ref{sec:main-res-exp}, and proved in Subsection~\ref{sec:proof-theor-refthrm}. The second scheme is introduce Subsection~\ref{subsec:concatenatedReliability}.
A concluding discussion appears in Subsection~\ref{exp-discussion}.

\subsection{Block Codes and Error Exponents}\label{sec:block-codes-error}
In what follows, we replace the scalar PAM mapping of the message point $W\rightarrow \Theta$ with an AWGN channel block code mapping $W\rightarrow \bM{\Theta}$ of length $N$. In this subsection we cite the classic results on the performance of block-codes for the AWGN channel. For channel coding over the AWGN channel with a signal-to-noise ratio $\snr$ and rate $R$, there exist block codes of length $N$ whose average error probability (averaged over the messages) under maximum likelihood decoding is exponentially upper bounded by $\Pr(\wh{W}(\bM{Y})\neq W) \dotleq e^{-N E_r(\snr,R)}$ where $E_r(\snr,R)$ is given by \cite{GallagerIT}:
\begin{align}
E_r(\snr,R) = 
\begin{cases}
E_{sp}(\snr,R) & \text{if } R_{rc}<R\leq C  \\
E_{rc}(\snr,R) & \text{if } R_{ex}<R\leq R_{rc} \\
E_{ex}(\snr,R) & \text{if } 0 < R \leq R_{ex} \\
\end{cases}
\end{align}
The boundaries between the regions are as follows. The \textit{Shannon capacity} is $C\dfn\frac{1}{2}\log(1+\snr)$. The \textit{critical rate} is $R_{cr}\dfn \sfrac{1}{2}\log\left(\sfrac{1}{2}+\sfrac{\snr}{4}+\sfrac{1}{2}\sqrt{1+\sfrac{\snr^2}{4}}\right)$. The \textit{expurgation rate} is $R_{ex}\dfn \sfrac{1}{2}\log\left(\sfrac{1}{2}+\sfrac{1}{2}\sqrt{1+\sfrac{\snr^2}{4}}\right)$. The exponents in the above three regions are given by:
\begin{align}
\label{eq:Esp}
E_{sp}(\snr,R)&=\tfrac{\snr}{4\beta}\left(\beta+1-(\beta-1)\sqrt{1+\tfrac{4\beta}{\snr(\beta-1)}}\right)\nonumber \\
&+\tfrac{1}{2}\ln\left(\beta- \tfrac{\snr(\beta-1)}{2}\sqrt{1+\tfrac{4\beta}{\snr(\beta-1)}}\right)
\end{align}
where $\beta=2^{2R}$, 
\begin{align}
E_{rc}(\snr,R)&=1-\beta+\frac{\snr}{2}+\frac{1}{2}\log\left(\beta-\frac{\snr}{2}\right)\\
&-\frac{1}{2}\log(\beta)-\log(2)R
\end{align}
where here $\beta=2e^{2R_{cr}}$, and 
\begin{align}
E_{ex}(\snr,R)&=\frac{\snr}{4}\left[1-\sqrt{1-2^{-2R}} \right].
\end{align}

It is also well known and readily verified that for $0<R<C$, the exponent $E_{sp}(\snr,R)$ coincides with the asymptotic expression of Shannon's \textit{sphere packing bound} for the AWGN channel~\cite{ShannonSpherePacking59}. Hence, $E_{sp}(\snr,R)$ is also an upper bound for the reliability function, and is tight above the critical rate. 

\subsection{JSCC with Side Information Using High Dimensional Lattices}\label{sec:jscc-with-side}

From this point on, boldface letters such as $\bM{X},\bM{\Theta}$ denote vectors of size $\latticeN$. As shown in 
Subsection~\ref{sec:joint-source-channel}, the probability of modulo-aliasing error in the JSCC problem with side information in its scalar version, while sometimes small, is bounded away from zero. In order to make this probability arbitrarily small, the dimension of the scheme should be increased. This can be achieved by introducing large dimensional lattices and replacing the scalar modulo with the corresponding modulo-lattice operations. Let us start by quickly surveying a few basic lattice notation and properties \cite{RamiLattices}:
\begin{enumerate}[(i)]
\item A lattice of dimension $\latticeN$ is denoted $\Lambda=G\cdot\mathbb{Z}^\latticeN$, where $G$ is the generating matrix. 
\item $\textrm{Vol}(\Lambda)=|\det(G)|$ is the lattice cell volume.
\item The nearest neighbor quantization of $\bM{x}$  w.r.t. $\Lambda$ is denoted $\quantLattice{\bM{x}}$.
\item $\V0=\{\bM{x}:\quantLattice{\bM{x}}=\bM{0} \}$ is the fundamental Voronoi cell pertaining to $\Lambda$. 
\item The modulo $\Lambda$ operation is $\moduloLattice{\bM{x}}\dfn \bM{x}-\quantLattice{\bM{x
}}$.
\item $\moduloLattice{\cdot}$ satisfies the \textit{distributive law}
  \begin{align}
    \moduloLattice{\moduloLattice{\bM{x}}+\bM{y}}=\moduloLattice{\bM{x}+\bM{y}}
  \end{align}
\item The volume to noise ratio (VNR) of a lattice in the presence of AWGN with variance $\sigma^2$ is $\mu(\Lambda)\dfn \left[\textrm{Vol}(\Lambda)\right]^{2/\latticeN}/\sigma^2$. 
\item The normalized second moment of a lattice $\Lambda$ is  $G(\Lambda)\dfn \sigma^2(\Lambda)/\left[\textrm{Vol}(\Lambda)\right]^{2/\latticeN}$, where $\sigma^2(\Lambda)=\frac{1}{\latticeN}\Expt{(\|\bM{V} \|^2)}$ and $\bM{V}$ is uniformly distributed on $\V0$.
\end{enumerate}

Consider again the JSCC problem introduced in Subsection~\ref{sec:joint-source-channel}, where now Terminal B is in possession of a vector $\wh{\bM{\Theta}}=\bM{\Theta}+\bM{\eps}$, and wants to convey the i.i.d. $\sim \mathcal{N}(0,\sigma^2_\eps)$ error vector $\bM{\eps}$ to Terminal A which is in possession of $\bM{\Theta}$, over an AWGN channel $\wt{Y}=\wt{X}+\wt{Z}$. The channel is again characterized by $\wt{Z}\sim \mathcal{N}(0,\wt{\sigma}^2)$ and $\Expt{\wt{X}^2}\leq \wt{P}$. We assume that a \textit{dither} signal  $\bM{V}\sim\textrm{Uniform}\left(\V0\right)$, mutually independent of the message and the channel noises, is known at both terminals. 

Let us revise the JSCC with side information scheme presented in Subsection~\ref{sec:joint-source-channel}, replacing the scalar modulo operation $\moduloOp{\cdot}$ with the lattice modulo operation $\moduloLattice{\cdot}$. Hence, Terminal B transmits 
\begin{align}
  \wt{X} = \moduloLattice{\gamma\wh{\bM{\Theta}}+\bM{V}},
\end{align}
Terminal A estimates:
\begin{align}
  \wh{\bM{\eps}}=\frac{1}{\gamma}\moduloLattice{\wt{\bM{Y}}-\gamma\bM{\Theta}-\bM{V}}=\frac{1}{\gamma}\moduloLattice{\gamma\bM{\eps}+\wt{\bM{Z}}}
\end{align}
And if $\gamma\bM{\eps}+\wt{\bM{Z}}\in \V0$ then
\begin{align}
  \wh{\bM{\eps}}=\bM{\eps}+\frac{1}{\gamma}\wt{\bM{Z}}
\end{align}
 
We are now ready to set the parameters of the modulo-lattice scheme. Let us set the lattice's second moment to equal the feedback power constraint $\sigma^2(\Lambda)=\wt{P}$. This would guarantee (due to dithering) that the feedback transmission power constraint is satisfied. The modulo-aliasing error event is the event where
\begin{align}
\label{eq:moduloError}
\gamma \bM{\eps} + \wt{\bM{Z}}\notin \V0
\end{align}
Recall the definition of the \textit{looseness} parameter $\latticeLoose$ of the lattice in~\eqref{eq:latticecLooseDef}, and note that $\latticeLoose = \mu(\Lambda)\cdot G(\Lambda)$. It was shown in \cite[Theorem 5]{GoodLattices} that there exist lattices that asymptotically attain $G(\Lambda)= \frac{1}{2\pi e} + o(1)$, 
and a modulo-error that is exponentially bounded by $\Pmod \dotleq e^{-\latticeN E_p\left(\frac{\mu(\Lambda)}{2\pi e}{}\right)}$, where $E_p(x)$ is the \textit{Poltyrev exponent}, given by \cite{RamiLattices}
\begin{align}
\label{eq:PoltyrevExp}
E_p(x) = 
\begin{cases}
\frac{1}{2}\left(x-1-\ln (x)\right) & \text{if } 1<x\leq 2 \\
\frac{1}{2}\left(\ln(x)+\ln(\frac{e}{4}) \right) & \text{if } 2<x\leq 4 \\
\frac{1}{8}x & \text{if } x>4 \\
\end{cases}
\end{align}
Hence in our notation, such lattices satisfy
\begin{align}
\label{eq:PmodLattice}
\Pmod \dotleq e^{-\latticeN E_p(L)}.
\end{align}

Clearly, in a similar fashion to the scalar scheme, the setting of  $\latticeLoose$ determines the trade-off between modulo-error probability and decision error probability. Setting $\latticeLoose$ close to $1$ will maximize the effective signal-to-noise ratio (and maximize the block code error exponent), but at the same time minimize the modulo-error exponent. Setting $\latticeLoose$ to be large will do the opposite, and will also reduce the maximal achievable rate. We note that the corresponding lattice-based JSCC scheme in \cite{KochmanZamirJointWZWDP} is better at low $\snr$, due to the addition of another Wiener coefficient multiplier at the receiver before the modulo operation, which results in non-Gaussian statistics of the error. For simplicity of analysis, this technique is not used here.

\subsection{Description of the Scheme}\label{sec:description-scheme}
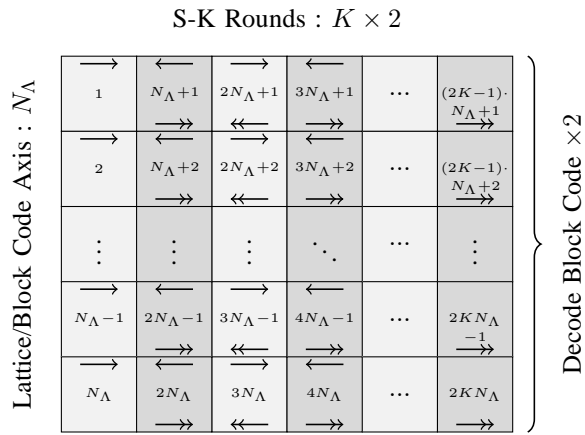
\begin{figure}
\centering
\newif\ifTIKZstandalone
\ifTIKZstandalone
\documentclass[10pt,journal,twocolumn]{IEEEtran}
\usepackage{amsmath}
\usepackage{amssymb,mathrsfs,dsfont}
\usepackage{tikz}
\usetikzlibrary{dsp,chains}
\usetikzlibrary{shapes,arrows}
\usetikzlibrary{positioning,shapes,shadows,arrows,scopes,decorations}
\usetikzlibrary{shapes.multipart}
\usetikzlibrary{snakes}
\def\wt{\widetilde}
\def\wh{\widehat}
\newcommand{\snr}{\mathrm{SNR}}
\newcommand{\moduloLattice}[1] {\mathbb{M}_{\Lambda}\left[#1\right]}
\newcommand{\bM}[1]{\boldsymbol{#1}}
\begin{document}
\else
\fi
\def\tikzuncover#1{}

\begin{tikzpicture}

\newcommand{\eval}[1]{\pgfmathparse{#1}\pgfmathresult}

\def\Ncode{{N_\Lambda}}
\definecolor{Gray1}{gray}{0.95}
\definecolor{Gray2}{gray}{0.85}
\newcommand{\smallfont}[1]{{\scriptscriptstyle #1}}

\def\xylab{{"$\smallfont{1}$","$\smallfont{\Ncode+1}$","$\smallfont{2\Ncode+1}$","$\smallfont{3\Ncode+1}$","$...$","$\smallfont{(2K-1)\cdot}$",
"$\smallfont{2}$","$\smallfont{\Ncode+2}$","$\smallfont{2\Ncode+2}$","$\smallfont{3\Ncode+2}$","$...$","$\smallfont{(2K-1)\cdot}$",
"$\vdots$","$\vdots$","$\vdots$","$\ddots$","$...$","$\vdots$",
"$\smallfont{\Ncode-1}$","$\smallfont{2\Ncode-1}$","$\smallfont{3\Ncode-1}$","$\smallfont{4\Ncode-1}$","$...$","$\smallfont{2K\Ncode}$",
"$\smallfont{\Ncode}$","$\smallfont{2\Ncode}$","$\smallfont{3\Ncode}$","$\smallfont{4\Ncode}$","$...$","$\smallfont{2K\Ncode}$"}};

\foreach \ii in {0,2,4}
{
\foreach \jj in {0,...,4}
{
  \draw (\ii,-\jj) +(-0.5,-0.5) rectangle ++(0.5,0.5) [fill=Gray1];
  \draw (\ii,-\jj) node {\eval{\xylab[\ii+6*\jj]}};
}
}

\foreach \ii in {1,3,5}
{
\foreach \jj in {0,...,4}
{
  \draw (\ii,-\jj) +(-0.5,-0.5) rectangle ++(0.5,0.5) [fill=Gray2];
  \draw (\ii,-\jj) node {\eval{\xylab[\ii+6*\jj]}};
}
}

\draw (5,-0.25) node {$\smallfont{ \Ncode+1}$};
\draw (5,-1.25) node {$\smallfont{ \Ncode+2}$};
\draw (5,-3.25) node {$\smallfont{-1}$};

\foreach \jj in {0,1,3,4}
{

\tikzuncover{1-}{
\def\ii{1}
\draw[->,semithick] (\ii-0.25-1,-\jj+0.4)--(\ii+0.25-1,-\jj+0.4);
}

\tikzuncover{1-}{
\def\ii{3}
\draw[<-,semithick] (\ii-0.25-2,-\jj+0.4)--(\ii+0.25-2,-\jj+0.4);
}

\tikzuncover{1-}{
\def\ii{3}
\draw[->,semithick] (\ii-0.25-1,-\jj+0.4)--(\ii+0.25-1,-\jj+0.4);
}

\tikzuncover{1-}{
\def\ii{5}
\draw[<-,semithick] (\ii-0.25-2,-\jj+0.4)--(\ii+0.25-2,-\jj+0.4);
}

\tikzuncover{2-}{
\def\ii{1}
\draw[->>,semithick] (\ii-0.25,-\jj-0.4)--(\ii+0.25,-\jj-0.4);
}

\tikzuncover{2-}{
\def\ii{4}
\draw[<<-,semithick] (\ii-0.25-2,-\jj-0.4)--(\ii+0.25-2,-\jj-0.4);
}

\tikzuncover{2-}{
\def\ii{3}
\draw[->>,semithick] (\ii-0.25,-\jj-0.4)--(\ii+0.25,-\jj-0.4);
}

\tikzuncover{2-}{
\def\ii{5}
\draw[->>,semithick] (\ii-0.25,-\jj-0.4)--(\ii+0.25,-\jj-0.4);
}

} 

\draw [] (2.5,1) node {S-K Rounds : $K\times 2$};
\draw [] (-1,-2.0) node [rotate=90] {Lattice/Block Code Axis : $\Ncode$ };
\tikzuncover{2-}
{
\draw [text width=4cm,text centered] (6.3,-2.0) node [rotate=90] {{Decode Block Code $\times 2$}};
\draw [semithick,snake=brace,mirror snake,segment amplitude = 5pt] (5.7,-4.5) -- ++(0,5);
}


\end{tikzpicture}

\ifTIKZstandalone
\end{document}
\fi
\caption{\label{fig:blockwiseTX} Blockwise transmission.
The time instants are divided into blocks of size $\Ncode$. Single headed arrows ``$\rightarrow$'' and ``$\leftarrow$'' denote transmission from Terminal A to Terminal B respectfully. Double headed errors , ``$\twoheadrightarrow$'' and ``$\twoheadleftarrow$'', bear the same meaning but for the second scheme.}
\end{figure}

In this subsection, we show how to combine the blockwise coding and blockwise modulo operations into one scheme. In a nutshell, the message $W$ is mapped into a codeword $\bM{\Theta}$ of length $\Ncode$, and sent in the first block ($\Ncode$ channel uses). This replaces the PAM transmission in the scalar scheme. In the sequel, vector analog transmission is used over the feedforward, and vector modulo-lattice transmission is used over the feedback. Ultimately, $W$ is decoded using a maximum likelihood decoding rule. Since under this protocol both terminals are idle half the time, we interlace two identical schemes, encoding and decoding two independent messages, as illustrated in Fig.~\ref{fig:blockwiseTX}. We denote the block index (or round index) by $k\in\{1,...,K\}$. For brevity, and with a mild abuse of notation, we only describe the evolution of one of the interlaced schemes. 

The setting of the parameters $\alpha,\beta_{\SKn},\gamma_{\SKn}$ will be discussed in the sequel. The dither variables $\bM{V}_{\SKn}$ are i.i.d., uniformly distributed on $\V0$, and mutually independent of the message and the noise processes.
\begin{enumerate}[(A)]
\item Initialization:
\begin{enumerate}[]
\item \textbf{Terminal A:} Map the message $W$ to codeword $\bM{\Theta}$ using a codebook for the AWGN channel with average power $P$. 
\item \textbf{Terminal A $\Rightarrow$ Terminal B:} 
  \begin{itemize}
  \item Send $\bM{X}_1=\bM{\Theta}$
    \item Receive $\bM{Y}_1=\bM{X}_1+\bM{Z}_1$
  \end{itemize}  

\item \textbf{Terminal B:} Initialize the $\bM{\Theta}$ estimate
to $\wh{\bM{\Theta}}_1=\bM{Y}_1$. 
\end{enumerate}
\item Iteration:
\begin{enumerate}[]
\item \textbf{Terminal B $\Rightarrow$ Terminal A:} 
  \begin{itemize}
  \item Given the $\bM{\Theta}$ estimate $\wh{\bM{\Theta}}_{\SKn}$, compute and send in the following block
    \begin{align}
      \wt{\bM{X}}_{\SKn}= \moduloLattice{\gamma_{\SKn}\wh{\bM{\Theta}}_{\SKn}
+\bM{V}_{\SKn}}
    \end{align}
  \item Receive $\wt{\bM{Y}}_{\SKn}=\wt{\bM{X}}_{\SKn} + \wt{\bM{Z}}_{\SKn}$
  \end{itemize}  
\item \textbf{Terminal A:} Extract a noisy scaled version of
the estimation error vector $\bM{\eps}_{\SKn}$:
\begin{align}
\wt{\bM{\eps}}_{\SKn}=\frac{1}{\gamma_k}\moduloLattice{\wt{\bM{Y}}_{\SKn}-\gamma_{\SKn}{\bM{\Theta}}-\bM{V}_{\SKn}} 
\end{align}
Note that $\wt{\bM{\eps}}_{\SKn}  =  \bM{\eps}_{\SKn} + \frac{1}{\gamma_k}\wt{\bM{Z}}_{\SKn}$, unless a modulo-aliasing error occurs. 
 \item \textbf{Terminal A $\Rightarrow$ Terminal B:} 
  \begin{itemize}
  \item Send a scaled version of $\wt{\bM{\eps}}_{\SKn}$:
$\bM{X}_{\SKn+1}=\alpha \gamma_k \wt{\bM{\eps}}_{\SKn}$, where $\alpha$ is set so that the input power constraint $P$ is met. 
  \item Receive $\bM{Y}_{\SKn+1}=\bM{X}_{\SKn+1}+\bM{Z}_{\SKn+1}$
  \end{itemize}  
\item \textbf{Terminal B:} 
Update the $\bM{\Theta}$ estimate
 $\wh{\bM{\Theta}}_{\SKn+1}=\wh{\bM{\Theta}}_{\SKn}-\wh{\bM{\eps}}_{\SKn}$, where  
\begin{align}
\label{eq:etahat}
\wh{\bM{\eps}}_{\SKn}=\beta_{\SKn+1}\bM{Y}_{\SKn+1}  
\end{align}
\end{enumerate}
\item Decoding: 
After the reception of block $\SKN$ the receiver decodes the message $\wh{W}(\wh{\bM{\Theta}}_{\SKN})$ using an ML decision rule w.r.t. the codebook. 
\end{enumerate}

\subsection{Main Result: The Error Exponent}\label{sec:main-res-exp}
Set the scheme parameters $\alpha,\beta_\SKn,\gamma_\SKn$ to
\begin{align}
\alpha=\sqrt{\latticeLoose\frac{P}{\wt{P}}},
\end{align}
\begin{align} 
\beta_{\SKn}=\frac{\sigma_{\SKn-1}}{\sigma}\frac{\sqrt{\snr\cdot\left( 1-\latticeLoose\bsnr^{-1}\right)}{ }}{1+\snr},
\end{align}
\begin{align}
  \gamma_\SKn = \frac{1}{\sigma_\SKn}\sqrt{\frac{\wt{P}}{\latticeLoose}-\wt{\sigma}^2}, 
\end{align}
where $\sigma_\SKn=\sigma_\SKn(\latticeLoose)=\frac{1}{\snr_\SKN(\latticeLoose)}$, and 
\begin{align}
\label{eq:SNRK}
\snr_\SKN(\latticeLoose)\dfn \snr\left(1+\snr\cdot\tfrac{1-\latticeLoose\bsnr^{-1}}{1+\latticeLoose\dsnr^{-1}}\right)^{\SKN-1},
\end{align}
The following theorem provides a lower bound on the error exponent obtained by our scheme. 
\begin{theorem}\label{thrm:exp}
For the choice of parameters above, the interactive communication scheme described in Subsection~\ref{sec:description-scheme} with a total delay of $N$ time instants, attains an error probability $\Pe \dotleq e^{-N E_\textrm{FB}(R)}$, where
\begin{align}
\label{eq:EFB}
E_\textrm{FB}(R)&\dfn\max_{\SKN\in\mathbb{N},L\geq 1} \left\{\frac{\min \left\{ E_r ( \snr_{\SKN}(\latticeLoose),\SKN R ),E_p(\latticeLoose) \right\}}{2\SKN}\right\}.
\end{align}
\end{theorem}

\subsection{Proof of Theorem \ref{thrm:exp}}\label{sec:proof-theor-refthrm}
Define the error event 
\begin{align}
E_{\SKN}\dfn \left\{\wh{W}(\wh{\bM{\Theta}}_{\SKN}) \neq W\right\}
\end{align}
The error probability of each of the interlaced schemes is $\Pe = \Pr(E_{\SKN})$, and hence the total error probability is upper bounded by $2\Pe$. Therefore, below we analyze only a single scheme, since the $2$ factor does not change the exponential behavior. As in the analysis of the scalar scheme, the channel $\bs{\Theta}\rightarrow \bM{Y}_K$ is not Gaussian due to the non-linear modulo operations, which complicates a direct analysis. In order to circumvent this, we will upper bound the error probability by further taking modulo-error events into account, and working in the couples system as before. These errors events are defined by 
\begin{align}
E_{\SKn}\dfn \left\{\gamma_\SKn \bM{\eps}_\SKn + \wt{\bM{Z}}_\SKn\notin \V0\right\}
\end{align}
and we have that 
\begin{align}
\Pe \leq \Pr\left(\bigcup_{\SKn=1}^\SKN E_{\SKn}\right).
\end{align}
Applying the coupling argument of Lemma~\ref{lemma:coupling}, we can obtain 
\begin{align}
\Pr\left(\bigcup_{\SKn=1}^\SKN E_\SKn \right) =  \Pr\left(\bigcup_{\SKn=1}^\SKN E_\SKn' \right).
\end{align}
Using the union bound we obtain
\begin{align}
\Pe\leq\sum_{\SKn=1}^{\SKN}\Pr\left( E_\SKn'\right).
\end{align}
Calculating the above probabilities now involves only Gaussian random vectors, which significantly simplifies the analysis.

From this point on, we perform an asymptotic exponential analysis. We set the parameters such that all modulo-aliasing error probabilities are equal. Hence
\begin{align}
\Pe \leq \sum_{k=1}^K \Pr(E_k') \dotleq \max(\Pmod, \Pdec)
\end{align}
where without loss of asymptotic optimality we have set all the modulo-error probabilities to be equal $\Pmod\dfn\Pr\left( E_k'\right)$, and also defined $\Pdec\dfn\Pr\left( E_\SKN'\right)$. The modulo-aliasing error can be exponentially upper bounded by the Poltyrev exponent \eqref{eq:PmodLattice}, i.e. $\Pmod \dotleq e^{-\latticeN E_p(L)}$.

We observe that the channel $\bM{\Theta}\to \bM{\wh{\Theta}}_\SKN$ is in fact equivalent (in the coupled system) to $\latticeN$ parallel independent AWGN channels each with the same noise variance $\sigma_{\SKN}^2$, and with a signal-to-noise ratio $\snr_\SKN(\latticeLoose)$ given in~\eqref{eq:SNRK}. We can now encode the message $W$ into $\bM{\Theta}$ using a Gaussian codebook of block length $\latticeN$ and rate $\SKN R$ to obtain 
\begin{align}
\Pdec\dotleq e^{-\latticeN  E_r\left(\snr_\SKN(\latticeLoose),\SKN R\right)}.
\end{align}
Note that the rate $\SKN R$ is chosen such that the overall rate over $K$ rounds is $R$. Balancing the exponents for $\Pdec$ and $\Pmod$ yields the result. The division by $2\SKN$ is due to the use of two interlaced schemes, that doubles the overall delay. 

The trade-off is now clear: setting the lattice looseness $\latticeLoose$ to be large reduces $\Pmod$ but also reduces $\snr_{\SKN}(L)$ hence increasing $\Pdec$, and vice versa. Due to the monotonicity of $E_r (\snr_{\SKN}(\latticeLoose),\SKN R ),E_p(\latticeLoose)$ in $\latticeLoose$, a numerical solution to \eqref{eq:EFB} can be easily found.  

\begin{remark}
It should be noted that as in the simple interaction case, the feedforward transmission power should be reduced by a suitable factor $\psid$ in order to avoid excess power due to modulo-aliasing errors. However, for high-dimensional lattices, the probability of a modulo-aliasing error is exponentially small in the dimension of the lattice, and the magnitude of the vectors in the fundamental Voronoi cell $\V0$ is upper bounded by $P$ \cite[Section~VIII]{GoodLattices}. Therefore, the power back-off term $\psid$ is asymptotically negligible and does not affect the statement of the theorem.  
\end{remark}

\subsection{Concatenated Coding with Simple Interaction \label{subsec:concatenatedReliability}}
In this subsection we introduce a concatenated coding scheme based on the simple interaction scheme of Section~\ref{sec:simplicity} used in conjunction with a (non-interactive) block code. The simple interaction scheme is used as an in inner code, inducing a DMC whose input and output correspond to a PAM constellation, and a suitable block code is employed as an outer code over that induced DMC. If the transition matrix pertaining to the induced DMC had been known, the calculation of the error exponent would have been straightforward. Unfortunately, computing this transition matrix appears to be extremely involved. Nevertheless, there is one  property of the transition matrix that is already available to us from the preceding discussion - an upper bound on the error probability, namely the probability that the output differs from the input. Therefore, to circumvent the difficulty in computing the error exponent of the induced DMC, we bound it from below by computing the error exponent corresponding to a ``worst case'' symmetric DMC, that has an error probability equal to (the upper bound on) the error probability induced by our inner coding scheme. This leads to the following result:

\begin{theorem}
	\label{theorem:concatenatedExp}
	The concatenated coding scheme described above, with total delay of $N$ time instants, attains an error probability $\Pe \dotleq e^{-N E^\textrm{DMC}_\textrm{FB}(R)}$, where 
	\begin{align}
	\label{eq:EFBHD}
	E^\textrm{DMC}_\textrm{FB}(R)&\dfn\max_{\PeDMC\in(0,\frac{1}{2}),K\in\mathbb{N}} 
	\left\{
	\frac{ E^\textrm{DMC} (\PeDMC,M,KR )}{K}
	\right\}. 
	\end{align}
	$M = \left\lfloor\sqrt{1+\frac{\snr_K}{\Gamma_0\left(\frac{\PeDMC}{2}\right)}}\right\rfloor$ is the size of the PAM constellation, $\snr_K$ is given in \eqref{eq:snreqmod} and  
	\begin{align}
	E^\textrm{DMC} (\PeDMC,M,R ) = \max
	\left\{
	E^\textrm{DMC}_r (\PeDMC,M,R ),
	E^\textrm{DMC}_{ex} (\PeDMC,M,R )
	\right\}.
	\end{align}
	The random coding error exponent is given by
	\begin{align}
	\label{eq:ErHD}
	&E^\textrm{DMC}_{r} (\PeDMC,M,R ) = \\
	&
	\max_{0\leq\rho\leq 1}
	\bigg\{
	-\rho\ln(2) R +\rho\ln(M)
	-\\
	&
	(1+\rho)\ln\left(
	(1-\PeDMC)^{\frac{1}{1+\rho}}
	+(M-1)^{\frac{\rho}{1+\rho}}
	\PeDMC^{\frac{1}{1+\rho}}
	\right)
	\bigg\}
	\end{align}
	and the expurgated error exponent is given by
	\begin{align}
	\label{eq:EexHD}
	&E^\textrm{DMC}_{ex} (\PeDMC,M,R ) = \\
	&\sup_{\rho\geq 1}\Bigg\{
	-\rho\ln(2) R+\rho\ln(M)
	\\&-\rho
	\ln\left(1+(M-1)
	\left(2\sqrt{\tfrac{\PeDMC(1-\PeDMC)}{M-1}}
	+\frac{M-2}{M-1}\PeDMC
	\right)^{\frac{1}{\rho}}
	\right)
	\Bigg\}.
	\end{align}	

\end{theorem}

\vspace{5pt}

Let $P$ denote an $M\times M$ transition matrix of a DMC, where $P\left(i\mid j\right)$ is the probability that the output is $j$ given that the input is $i$. The maximal error probability of $P$ is defined as 
\begin{align}
\maxerr(P) \dfn 1-\min_{i}P(i\mid i)  
\end{align}
We say that $P$ is \textit{totally symmetric}, if  
	\begin{align}
	\label{eq:PbarVals}
	P\left(i\mid j\right)=
	\begin{cases}
	1-\PeDMC & \text{if } i=j  \\
	\frac{\PeDMC}{M-1} & \text{if } i\neq j
	\end{cases}.
	\end{align}
For some $\PeDMC$. Clearly, for such $P$ it holds that $\maxerr(P) = \PeDMC$. 

The proof of the theorem is based on the observation that a totally symmetric DMC has the smallest uniform input random coding / expurgated error exponent for a given maximal error probability, together with straightforward calculations of the classic error exponents \cite{GallagerIT} for this channel. Specifically, denote the random coding error exponent at rate $R$ for a DMC with transition matrix $P$ and uniform input distribution by $E_r(R,P)$. The corresponding expurgated error exponent is denoted by $E_{ex}(R,P)$.  The worst case property of the totally symmetric DMC is stated in the following lemma, which is proved in Appendix~\ref{appendixA}:
	\begin{lemma}
	\label{lemma:worstcaseDMC}
	Let $P$ be a $M\times M$ DMC with $\maxerr(P)\leq \PeDMC$, and let $P_{sym}$ be a totally symmetric $M\times M$ DMC with $\maxerr(P_{sym}) = \PeDMC$. Then $E_r(R,P_{sym})\leq E_r(R,P)$ and $E_{ex}(R,P_{sym})\leq E_{ex}(R,P)$.
	\end{lemma}

\begin{IEEEproof}[Proof of Theorem~\ref{theorem:concatenatedExp}]
Let $P$ be the transition matrix pertaining to the $M\times M$ DMC induced by our simple interaction scheme, with some target error probability $\PeDMC$ and $K$ interaction rounds (the explicit derivation of the scheme parameters and the error analysis are given in Section~\ref{sec:simplicity}). This DMC satisfies $\maxerr(P) \leq \PeDMC$. Let $P_{sym}$ denote the transition matrix of a totally symmetric $M\times M$ DMC with $\maxerr(P_{sym}) = \PeDMC$. Lemma \ref{lemma:worstcaseDMC} implies that the error exponent attained by an optimal outer block code is lower bounded by the random coding / expurgated exponents of $P_{sym}$, where the latter follow from a straightforward calculation and are given by \eqref{eq:ErHD} and \eqref{eq:EexHD}. Finally, \eqref{eq:EFBHD} follows from the appropriate normalization in the inner code delay $K$, and by an optimization over the target error probability and the number of interaction rounds. 
\end{IEEEproof} 

\subsection{Discussion}
\label{exp-discussion}
\begin{figure}
\centering
\input{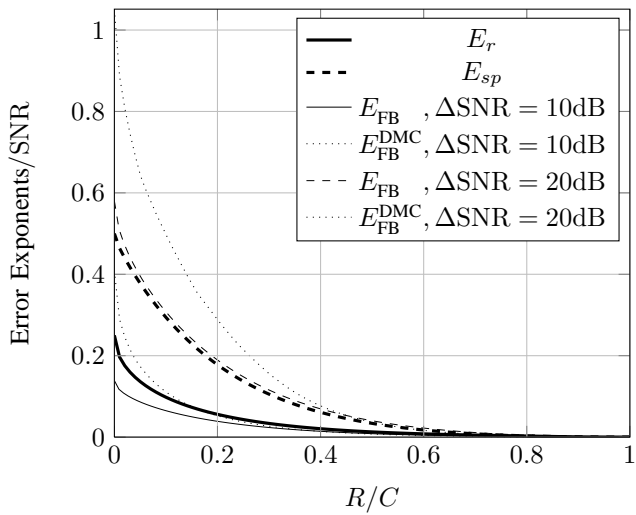}
\caption{\label{fig:EFB_snr20}
Error exponents with and without feedback for $\snr=20\db$ with $\dsnr=10\db$ and $\dsnr=20\db$}
\end{figure}

\begin{figure}
	\centering
	\input{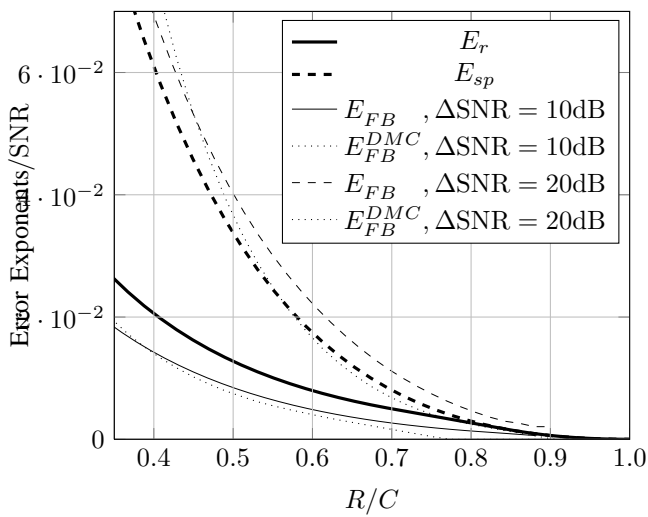}
	\caption{\label{fig:EFB_snr20zoomin}
		 Fig.~\ref{fig:EFB_snr20} at high rates}
\end{figure}

Numerical evaluations of  $E_{sp}$,  $E_{r}$, $E_{\textrm{FB}}$ and $E^{\textrm{DMC}}_{\textrm{FB}}$ for $\snr=20\db$ are depicted in Fig.~\ref{fig:EFB_snr20}. It can be clearly seen that for low rates
$E^{\textrm{DMC}}_{\textrm{FB}}$ is larger than $E_{\textrm{FB}}$ for both value of $\dsnr$. What is less visible in Fig.~\ref{fig:EFB_snr20} but is clearer in Fig.~\ref{fig:EFB_snr20zoomin} is that at high rates $E_{\textrm{FB}}$ exceeds $E^{\textrm{DMC}}_{\textrm{FB}}$. Comparing these achievable exponent to the sphere packing bound, which upper bounds the best achievable error exponent without feedback, it is evident that both error exponents are smaller than the sphere packing bound for $\dsnr=10\db$, and the maximal between them is greater than the sphere packing bound for $\dsnr = 20\db$. 

It is of interest to compare the error exponents of Theorems \ref{thrm:exp} and \ref{theorem:concatenatedExp} to the noisy feedback error exponent of \cite{ChanceLove}, which employs a concatenated linear coding scheme. In this scheme the output of the inner code is a linear combination of the message variable and noisy versions of the previous channel outputs (i.e. \textit{passive} feedback), and the outer code is a random block code. Due to the linear nature of the inner code, its output is regarded as a single use of an AWGN channel, and the scheme parameters are judiciously chosen to try and maximize its $\snr$. Then, the error exponent of the concatenated scheme is readily calculated using the standard AWGN error exponents.

\begin{figure}
\centering
\input{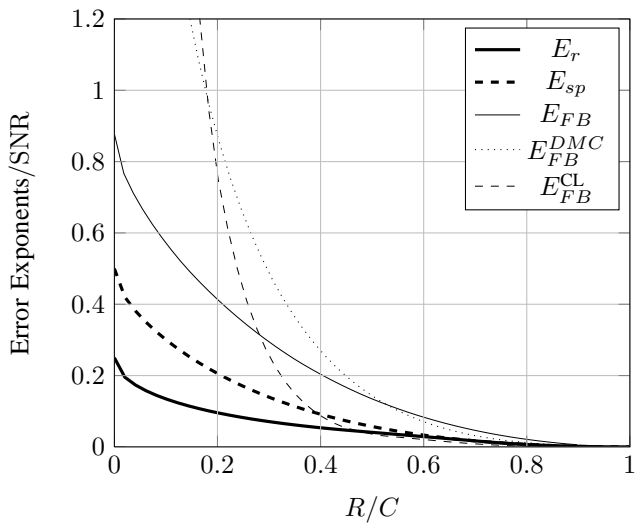}
\caption{
\label{fig:ChanceLove}
A comparison of non-feedback error exponents, our error exponents ($E_{FB},E_{FB}^{DMC}$) and Chance and Love's error exponent ($E_{FB}^{CL}$). In this setting $\snr=10\db$ and $\dsnr=23\db$
}
\end{figure}

Fig.~\ref{fig:ChanceLove} compares our error exponents to that of \cite{ChanceLove}, which is denoted by $E_{\text{FB}}^{CL}$. In this setting $\snr=10\db$ and $\dsnr=23\db$. We can see that $E_{FB}^{CL}$ dramatically improves over the non-feedback achievable error exponent at rates close to zero, but then falls below it at rates above $0.46C$ (where $C$ is the channel capacity). Comparing  $E_{\text{FB}}^{CL}$ to $E^{\textrm{DMC}}_{\textrm{FB}}$ we can see that the latter is better at rates higher than $0.18C$. Finally, 
$E_{\textrm{FB}}$ is the best among the three feedback error exponent at rates above $0.53C$, and also beats the sphere packing bound at rates up to $0.9C$. 

It is interesting to note that in \cite{ChanceLove}, the authors characterize the ``shut-off rate'' $R_{th}^{CL}$ of their scheme, namely as the rate above which their exponent does not improve the no-feedback one. This rate is given by 
\cite[Lemma 8]{ChanceLove}:
\begin{align}
R_{th}^{CL} = \frac{1}{4}\log\left(1+2\snr\cdot \dsnr\cdot \frac{\snr}{1+\snr}(1-\gamma_0)\right)
\end{align}
where $\gamma_0\in[0,1]$ is a root of a quadratic equation given in their Lemma 6 (note that a factor $\tfrac{1}{2}$ is missing in the original expression for $R_{th}^{CL}$). Dividing the above by the capacity, it is easy to show that 
\begin{align}
&\frac{R_{th}^{CL}}{C}  \leq\\&  \frac{1}{2}\left(1+\frac{1+\log(1+\dsnr)}{\log\snr} + \frac{(1+\snr)\log{e}}{2\dsnr\cdot \snr^2\log\snr}\right)
\end{align}
For $\snr \gg 1$ and a fixed $\dsnr$, the above upper bound clearly converges to $\tfrac{1}{2}$, hence in this high-SNR regime the error exponent of \cite{ChanceLove} does not yield any improvement over the non-feedback exponent for rates above half the capacity. 

It is also instructive to compare the error exponents discussed above at zero rate. The zero-rate exponent attained by the scheme in \cite{ChanceLove} is given in Lemma 8 therein:
\begin{align}
\label{eq:CLR0}
  E_{FB}^{CL}(R=0) &= \frac{\snr}{4}\left(1+\dsnr\cdot \frac{\snr}{1+\snr}\right) 
\end{align}

It can be verified by direct calculation that at zero rate, the error exponent of our concatenated scheme outperforms that of the lattice-based scheme. The former is given in the following formula at high-$\snr$, derived in Appendix~\ref{app:R0EE}:
\begin{align}
\label{eq:DMCR0}
&E^\textrm{DMC}_\textrm{FB}(R=0)\geq\\&
\frac{\frac{3}{2}\dsnr\left(e^{-1}\snr-1\right)-\ln(2\lfloor\ln(3\dsnr)\rfloor-1)}
{4\lfloor\ln(3\dsnr)\rfloor}
\end{align}
It is easy to see that in the limit of $\snr\rightarrow\infty$  \eqref{eq:CLR0} is largers than 
\eqref{eq:DMCR0}.

\section{Acknowledgments}
We thank the anonymous reviewers for many helpful comments and observations that have significantly improved the presentation of the paper. 

\appendices
\section{Proof Lemma~\ref{lemma:worstcaseDMC}\label{appendixA}}
The proof is based on a symmetrization argument in conjunction with a convexity argument. We begin by recalling the error exponents for a DMC with transition matrix $P$, input distribution $Q$ and rate $R$ \cite{GallagerIT}. The random coding exponent is given by
\begin{align}
&E_r(R,Q,P)=\\&\max_{0\leq \rho \leq 1}
\left\{
-\rho \ln(2) R-\ln\sum_{j=0}^{M-1}
\left[\sum_{k=0}^{M-1} 
Q(k)P(j\mid k)^{\frac{1}{1+\rho}}
\right]^{1+\rho} 	
\right\},
\end{align}
and the expurgated error exponent is given by
\begin{align}
&E_{ex}(R,Q,P) =\max_{\rho\geq 1}
\Bigg\{
-\rho \ln(2)R  \\&
-\rho\ln\sum_{k,i}Q(k)Q(i)\left[\sum_j\sqrt{P(j\mid k)P(j\mid i)}\right]^{\frac{1}{\rho}}
\Bigg\}.
\end{align}	

Below we restrict our attention to a uniform input distribution $Q(i)=\frac{1}{M}$ for $0\leq i\leq M-1$, which can be suboptimal but facilitates a much simpler analysis. We denote the random coding and expurgated exponents corresponding to a uniform input by $E_r(R,P)$ and $E_{ex}(R,P)$ respectively. These are explicitly given by 
\begin{align}
&E_r(R,P)=\\&
\max_{0\leq \rho \leq 1}
\left\{
-\rho \ln(2) R
+(1+\rho)\ln(M)-
\ln F_r(\rho,P)
\right\},\label{eq:ErDMC}
\end{align}			
and 
\begin{align}
&E_{ex}(R,P) =\\&\max_{\rho\geq 1}
\left\{
-\rho \ln(2)R  
+2\rho\ln(M)
-\rho\ln F_{ex}(\rho,P)
\right\},\label{eq:ExDMC}
\end{align}	
where
\begin{align}
F_r(\rho,P)\dfn \sum_{j=0}^{M-1}
\left[\sum_{k=0}^{M-1} 
P(j\mid k)^{\frac{1}{1+\rho}}
\right]^{1+\rho}
\end{align}
and 
\begin{align}
F_{ex}(\rho,P)\dfn\sum_{k,i}\left[\sum_j\sqrt{P(j\mid k)P(j\mid i)}\right]^{\frac{1}{\rho}}.
\end{align}

Now, let $\Pi$ denote the set of all possible permutations over $\{1,\ldots,M\}$. For a given channel transition matrix $P$, and a given permutation $\pi\in\Pi$, we define the channel transition matrix $P_\pi$ to be 
\begin{align}
  P_\pi(j \mid i) \dfn P(\pi(j)\mid \pi(i)).
\end{align}
The following lemma easily follows from \eqref{eq:ErDMC} and \eqref{eq:ExDMC} by incorporating the permutation into the order of summations. 
\begin{lemma}
	\label{lemma:permutation}
	$E_r(R,P_\pi)=E_r(R,P)$ and $E_{ex}(R,P_\pi)=E_{ex}(R,P)$ for any $\pi\in\Pi$. 
\end{lemma}

We further define the symmetrized channel matrix $P_{sym}$ associated with $P$, to be the one obtained by uniformly averaging over all possible permutations, i.e., 
\begin{align}
  P_{sym}(j \mid i) \dfn \frac{1}{|\Pi|}\sum_{\pi\in \Pi} P\pi(j\mid i).
\end{align}

We now have the following lemma. 
\begin{lemma}
  $P_{sym}$ is a totally symmetric DMC with $\maxerr(P_{sym}) \leq \maxerr(P)$. 
\end{lemma}
\begin{IEEEproof}
Choose some indices $i,j,k,\ell$, such that if $i=j$ then also $k=\ell$. Let $\sigma\in \Pi$ be the permutation where $\sigma(i)=k$,  $\sigma(j)=\ell$, and $\sigma(x)=x$ otherwise. 
  \begin{align}
  P_{sym}(j \mid i) &= \frac{1}{|\Pi|}\sum_{\pi\in \Pi} P(\pi(j)\mid \pi(i))\\
& = \frac{1}{|\Pi|}\sum_{\pi\in \Pi} P((\pi\circ\sigma) (j)\mid (\pi\circ\sigma)(i))\\
& = \frac{1}{|\Pi|}\sum_{\pi\in \Pi} P(\pi(k)\mid \pi(\ell))\\
& = P_{sym}(k \mid \ell).
\end{align}
Hence $P_{sym}$ is a totally symmetric DMC. Specifically, for any $i$ 
  \begin{align}
  P_{sym}(i \mid i) &= \frac{1}{|\Pi|}\sum_{\pi\in \Pi} P(\pi(i)\mid \pi(i))\\
& = \frac{1}{|\Pi|}\sum_{k}\sum_{\pi\in\Pi, \pi(i)=k} P(k\mid k)\\
& = \frac{1}{M}\sum_{k} P(k|k).
\end{align}
Thus, 
\begin{align}
\maxerr(P_{sym}) &=
 1-\frac{1}{M}\sum_{k} P(k|k)\\ &\leq 1-\min_k P(k|k) 
 \\&= \maxerr(P)  .
\end{align}
\end{IEEEproof}

We now proceed to show that the (uniform input) error exponents associated with $P_{sym}$ can be used as lower bounds for those of $P$. This follows from a concavity argument combined with Lemma~\ref{lemma:permutation}.
\begin{lemma}
$E_r(R,P_{sym})\leq E_r(R,P)$ and $E_{ex}(R,P_{sym})\leq E_{ex}(R,P)$. 
\end{lemma}
\begin{IEEEproof}
We begin by showing that both $F_{r}(\rho,P)$ (in \eqref{eq:ErDMC}) and $F_{ex}(\rho,P)$ (in \eqref{eq:ExDMC}) are concave in $P$ for all valid values of $\rho$. We start with $F_{r}(\rho,P)$ and recall its definition:
	\begin{align}
	F_r(\rho,P)= \sum_{j=0}^{M-1}
	\left[\sum_{k=0}^{M-1} 
	P(j\mid k)^{\frac{1}{1+\rho}}
	\right]^{1+\rho}.
	\end{align}
	We note that $\sum_{k=0}^{M-1} P(j\mid k)^{\frac{1}{1+\rho}}$ is a concave function in the vector $P(j\mid \cdot)$; this follows by observing that this function is a norm of order $\frac{1}{1+\rho}\in[\frac{1}{2},1]$ over $\RealF_+^M$  (and the values of $P$ are strictly positive by design). $F_r(\rho,P)$ is thus concave in $P$ as well, being a sum of concave functions. 
	
	Similarly, we recall the definition of $F_{ex}(\rho,P)$
	\begin{align}
	F_{ex}(\rho,P)\dfn\sum_{k,i}\left[\sum_j\sqrt{P(j\mid k)P(j\mid i)}\right]^{\frac{1}{\rho}}.
	\end{align}
	We note that $\sqrt{P(j\mid k)P(j\mid i)}$ is a geometric mean, which is concave in the vector $(P(j\mid k),P(j\mid i))\in \RealF_+^M$ for any fixed $k$. Thus, $\sum_j\sqrt{P(j\mid k)P(j\mid i)}$ is concave as well, being a sum of concave functions. The $\frac{1}{\rho}$ power of this sum is concave since it is a scalar composition of a concave function with a concave nondecreasing function $x^{\frac{1}{\rho}}$ (for $\rho\geq 1$). Finally, $F_{ex}(\rho,P)$ is a sum of concave functions hence is also concave. 
	
	With concavity in hand, we can readily apply Jensen's inequality
	\begin{align}
	F_{r}(\rho,P_{sym})=&F_{r}\left(\rho,\frac{1}{|\Pi|}\sum_{\pi\in \Pi} {P}_{\pi}\right)\\
	\geq& \frac{1}{|\Pi|}\sum_{\pi\in \Pi} F_r\left(\rho,P_{\pi}\right)\label{eq:concavity_arg}\\
	=&F_r\left(\rho,P\right) \label{eq:samechans}
	\end{align}
	where~\eqref{eq:concavity_arg} follows from concavity via Jensen's inequality, and~\eqref{eq:samechans} is by virtue of Lemma.~\ref{lemma:permutation}. Plugging the resulting inequality in \eqref{eq:ErDMC} yields $E_r(R,P_{sym})\leq E_r(R,P)$. Following the exact same steps for $F_{ex}$ we can prove that $E_{ex}(R,P_{sym})\leq E_{ex}(R,P)$.
\end{IEEEproof}

Let us now calculate the error exponents for $P_{sym}$. Since by Lemma~\ref{lemma:permutation} we have that $P_{sym}$ is a totally symmetric DMC, and denoting $\PeDMC \dfn \maxerr(P_{sym})$, we can use  \eqref{eq:ErDMC} and \eqref{eq:ExDMC} to obtain
\begin{align}
\label{eq:ErPbar}
E_{r} (R,P_{sym} ) &= 
\max_{0\leq\rho\leq 1}
\bigg\{
-\rho\ln(2) R +\rho\ln(M)
-\\&(1+\rho)\ln\left(
(1-\PeDMC)^{\frac{1}{1+\rho}}
+(M-1)^{\frac{\rho}{1+\rho}}
{\PeDMC}^{\frac{1}{1+\rho}}
\right)
\bigg\}
\end{align}
and 
\begin{align}
\label{eq:ExPbar}
&E_{ex} (R,P_{sym} ) = 
\sup_{\rho\geq 1}\Bigg\{
-\rho\ln(2) R+\rho\ln(M)
-\\&\rho\ln\left(1+(M-1)
\left(2\sqrt{\tfrac{\PeDMC(1-\PeDMC)}{M-1}}
+\frac{M-2}{M-1}\PeDMC
\right)^{\frac{1}{\rho}}
\right)
\Bigg\}.
\end{align}	

Lemma~\ref{lemma:permutation} also tells us that $\PeDMC \leq \maxerr{P}$. It can be verified (by direct differentiation) that both \eqref{eq:ErPbar} and \eqref{eq:ExPbar} are monotonically decreasing in $\PeDMC$ for $\PeDMC\in(0,\frac{1}{2})$. Therefore, replacing $\PeDMC$ with any upper bound on $\maxerr(P)$ still results in a lower bound for both $E_r(R,P_{sym})$ and $E_{ex}(R,P_{sym})$, concluding the proof.

\section{Zero Rate Analysis for Theorem~\ref{theorem:concatenatedExp}\label{app:R0EE}}
Below we provide a lower bound for the error exponent of Theorem~\ref{theorem:concatenatedExp} at $R=0$.  We use the expurgated error exponent which is known to be larger at $R=0$, and use $M=2$ in \eqref{eq:EexHD} which yields
\begin{align}
&E^\textrm{DMC}_{ex} (\PeDMC,M=2,R=0) = \\
&\sup_{\rho\geq 1}\left\{
\rho\ln(2)
-\rho
\ln\left(1+
\left(2\sqrt{\PeDMC(1-\PeDMC)}
\right)^{\frac{1}{\rho}}
\right)
\right\}.
\end{align}	
It can be shown (see \cite[Probelm 5.24]{GallagerIT}) that for every input distribution, the zero rate error exponent is optimized by taking the limit $\rho\rightarrow\infty$, which in the case of uniform input results in:
\begin{align}
E^\textrm{DMC}_{ex} (\PeDMC,M=2,R=0) 
&= -\frac{1}{2}\ln\left(2\sqrt{\PeDMC(1-\PeDMC)}\right)\\
&= -\frac{1}{4}\ln\left(2\PeDMC(1-\PeDMC)\right)\\
\label{eq:ExHDbound}
&\geq -\frac{1}{4}\ln(2\PeDMC)
\end{align}

Let us now find a simple setting for the scheme's parameters. Applying the union bound it is clear that the error probability of the inner code is upper bounded by $\PeDMC$, where
\begin{align}
\label{eq:PeBPSK}
\PeDMC=Q\left(\sqrt{\snr_K}\right)+2(K-1)Q\left(\sqrt{3L}\right),
\end{align}
where the first addend is due to the probability of error at the PAM (here, BPSK) decoder, and the second addend is due to the modulo-aliasing error probabilities. In order to simplify the solution, we impose equality between the two $Q$-functions, which results in $\snr_K=3L$. Using \eqref{eq:snreqmod} and some algebra, we obtain
\begin{align}
\label{eq:Lequation}
\snr\left(\frac{\dsnr(1+\snr)}{\dsnr+L}\right)^{K-1}=3L.
\end{align}
This equation is difficult to solve. To simplify, we note that solution of the following equation for $L$:
\begin{align}
\snr\left(\frac{\dsnr(1+\snr)}{\dsnr+L}\right)^{K-1}=3(\dsnr+L),
\end{align}
gives a lower bound on the corresponding solution of \eqref{eq:Lequation}, yielding
\begin{align}
&L=\\&-\dsnr+3^{-\frac{1}{K}}\dsnr^{1-\frac{1}{K}}\left(\snr(1+\snr)^{K-1}\right)^{\frac{1}{K}}.
\label{eq:Lexact}
\end{align}
It is instructive to use a high-$\snr$ approximation $1+\snr\approx\snr$ and compute an approximate (yet attainable) version of \eqref{eq:Lexact}:
\begin{align}
L=\dsnr\left(\left(3\dsnr\right)^{-\frac{1}{K}}\snr-1\right)
\end{align}
Plugging this into \eqref{eq:PeBPSK} and using the exponential bound for the Q-function: 
$Q(x)\leq\frac{1}{2}\exp(-\frac{1}{2}x^2)$, we get
\begin{align}
\PeDMC \leq (2K-1)Q(\sqrt{3L})\leq\frac{2K-1}{2}\exp\left(-\frac{3}{2}L\right)
\end{align}
Now, plugging the above into the error exponent bound \eqref{eq:ExHDbound} and recalling the concatenated coding exponent \eqref{eq:EFBHD}, results in
\begin{align}
&E^\textrm{DMC}_\textrm{FB}(R=0)\\ &\geq
\max_{K\in\mathbb{N}}
\left\{
\frac{\frac{3}{2}L-\ln(2K-1)}{4K}
\right\}\\
&=\label{eq:EHD0}
\max_{K\in\mathbb{N}}
\left\{
\frac{\frac{3}{2}\dsnr\left(\left(3\dsnr\right)^{-\frac{1}{K}}\snr-1\right)-\ln(2K-1)}{4K}
\right\}
\end{align}
Let us now approximately optimize for $K$, by taking the derivative to zero in a simplified expression
\begin{align}
\frac{(3\dsnr)^{-\frac{1}{K}}}{K}
\end{align}
yielding $K^*=\lfloor\ln(3\dsnr)\rfloor$ and $(3\dsnr)^{-\frac{1}{K^*}}\geq (3\dsnr)^{-\frac{1}{\ln(3\dsnr)}}=e^{-1}$ finally obtaining the following achievable error exponent:
\begin{align}
\label{eq:EfbDMCR0}
&E^\textrm{DMC}_\textrm{FB}(R=0)\geq\\&
\frac{\frac{3}{2}\dsnr\left(e^{-1}\snr-1\right)-\ln(2\lfloor\ln(3\dsnr)\rfloor-1)}
{4\lfloor\ln(3\dsnr)\rfloor}
\end{align}

Let us now evaluate $E^\textrm{DMC}_\textrm{FB}(R=0)$ for $\snr>>1$ when $\dsnr$ is held fixed. 
In this case \eqref{eq:EfbDMCR0} can be approximately expressed as
\begin{align}
\label{eq:EfbDMCR0highSNR}
E^\textrm{DMC}_\textrm{FB}(R=0)\gtrapprox
\frac{3\dsnr}{8 e\lfloor\ln(3\dsnr)\rfloor}\snr.
\end{align}

Note that by \eqref{eq:Esp} the sphere packing bound at $R=0$ is $\frac{1}{2}\snr$ rendering \eqref{eq:EfbDMCR0highSNR} better for $\dsnr_{\db}\gtrapprox 10.4\db$. It is interesting to note that in this setting \eqref{eq:CLR0} still yields a larger value.

\bibliographystyle{IEEEbib}
\bibliography{bibtex_references}

%

\begin{IEEEbiographynophoto}{Assaf Ben-Yishai}
received the B.Sc. degree (summa cum laude) and  M.Sc. degree (magna cum laude) in 1999 and 2001 respectively, both in electrical engineering, and both from the Tel-Aviv university. Assaf worked for several years in leading algorithmic positions in the Israeli high-tech industry, primarily in the design of advanced features of wireless and wire-line communication systems. Currently he is pursuing his Ph.D. degree in the field of information theory.
\end{IEEEbiographynophoto}

\begin{IEEEbiographynophoto}{Ofer Shayevitz }
received the B.Sc. degree (summa cum laude) from the
Technion Institute of Technology, Haifa, Israel, in 1997 and the M.Sc. and
Ph.D. degrees from the Tel-Aviv University, Tel Aviv, Israel, in 2004 and 2009,
respectively, all in electrical engineering. He is currently a Senior Lecturer in
the Department of EE - Systems at the Tel Aviv University, and also serves
as the head of the Advanced Communication Center (ACC). Before joining
the department, he was a postdoctoral fellow in the Information Theory and
Applications (ITA) Center at the University of California, San Diego, from
2008 to 2011, and worked as a quantitative analyst with the D. E. Shaw group
in New York from 2011 to 2013. Prior to his graduate studies, he served as
an engineer and team leader in the Israeli Defense Forces from 1997 to 2003,
and as an algorithms engineer at CellGuide from 2003 to 2004. Dr. Shayevitz
is the recipient of the ITA postdoctoral fellowship (2009 - 2011), the Adams
fellowship awarded by the Israel Academy of Sciences and Humanities
(2006 - 2008), the Advanced Communication Center (ACC) Feder Family
award (2009), and the Weinstein prize (2006 - 2009).
\end{IEEEbiographynophoto}





\end{document}